\documentclass[paper]{JHEP}
\usepackage{amsmath}
\usepackage{cite}
\usepackage{epsfig}

\def\lrang#1{\left\langle#1\right\rangle}
\def\abs#1{\left| \: #1 \: \right|}
\def\la{\mathrel{\mathpalette\fun <}}
\def\ga{\mathrel{\mathpalette\fun >}}
\def\fun#1#2{\lower3.6pt\vbox{\baselineskip0pt\lineskip.9pt
  \ialign{$\mathsurround=0pt#1\hfil##\hfil$\crcr#2\crcr\sim\crcr}}}
\def\cO#1{{\cal{O}}\left(#1\right)}

\def\vkti#1{\vec{k}_{t#1}}
\def\tkt{\kappa_t}
\def\vtkt{\vec{\kappa}_t}

\def\tkti#1{\kappa_{t#1}}
\def\vtkti#1{\vec{\kappa}_{t#1}}
\def\tpt{p}
\def\vtpt{\vec{\tpt}}

\def\vb{\vec{b}}

\def\bb{\bar{b}}

\def\al{\alpha}
\def\be{\beta}

\def\om{\omega}
\def\Om{\Omega}

\def\cA{{\cal{A}}}    
\def\cM{{\cal{M}}}    
\def\cI{{\cal{I}}}    %
\def\cN{\sigma}    %

\def\cR{{\cal{R}}}               

\def\half{\mbox{\small $\frac{1}{2}$}}
\def\tq{\mbox{\small $\frac{3}{4}$}}
\def\th{\mbox{\small $\frac{3}{2}$}}

\def\PT{\mbox{\scriptsize PT}}
\def\NP{\mbox{\scriptsize NP}}
\def\MSbar{\overline{\mbox{\scriptsize MS}}}

\def\as{\alpha_{\mbox{\scriptsize s}}}

\def\ae{\alpha_{\mbox{\scriptsize eff}}}

\def\LQCD{\Lambda_{\mbox{\scriptsize QCD}}}
\def\scale{\lambda}   

\def\EEC{{\rm EEC}}
\def\qq{{q\bar q}}
\def\bmin{b_{\mbox{\scriptsize min}}}
\def\bmax{b_{\mbox{\scriptsize max}}}
\def\stot{\sigma_{\mbox{\scriptsize tot}}}
\def\lms{\Lambda_{\overline{\mbox{\scriptsize MS}}}}
\def\mI{\mu_{\mbox{\scriptsize I}}}

\def\prep#1#2#3{{\em Phys. Rep.} {\bf {#1}} (19#3) #2}

\def\spj#1#2#3{{\em Sov. Phys. JETP} {\bf {#1}} (19#3) #2}

\def\zp#1#2#3{{\em Zeit.\ Phys.} {\bf {C #1}} (19#3) #2}

\def\epj#1#2#3{{\em Eur. Phys. J. } {\bf {C #1}} (19#3) #2}

\title{
Non-perturbative effects in\\
the energy-energy correlation}

\author{
Yu.L. \ Dokshitzer\thanks{On leave from St. Petersburg Nuclear Institute,
Gatchina, St. Petersburg 188350, Russia},
 G.\ Marchesini, \\
Dipartimento di Fisica, Universit\`a di Milano-Bicocca \\
and INFN, Sezione di Milano, Italy}
\author
{B.R. Webber\\
Cavendish Laboratory, University of Cambridge \\
Madingley Road, Cambridge CB3 0HE, UK }

\abstract{ The fully resummed next-to-leading-order perturbative
  calculation of the energy-energy correlation in $e^+e^-$
  annihilation is extended to include the leading non-perturbative
  power-behaved contributions computed using the ``dispersive method''
  applied earlier to event shape variables.  The correlation between a
  leading (anti)quark and a gluon produces a non-perturbative $1/Q$
  contribution, while non-perturbative effects in the quark-antiquark
  correlation give rise to a smaller contribution $\ln Q^2/Q^2$.  In
  the back-to-back region, the power-suppressed contributions actually
  decrease much more slowly, as small non-integer powers of $1/Q$, as
  a result of the interplay with perturbative effects.

The hypothesis of a universal low-energy
form for the strong coupling relates the coefficients of
these contributions to those measured for other observables.
}


\preprint{
     Bicocca--FT--99/01\\
     Cavendish--HEP--99/01\\
     hep-ph/9905339 \\
     May 1999 }

\begin{document}

\section{Introduction}
The energy-energy correlation \cite{BBEL} (EEC) in $e^+e^-$
annihilation was one of the first collinear and infrared safe 
observables \cite{SW} for which the all-order resummation of
perturbative (PT) radiative corrections proved to be necessary, in the
back-to-back (as well as in the forward) kinematical configuration
\cite{DDT,CS81,RW}. It was soon noticed that the comparison of the
theoretical prediction with the data required the introduction of
sizeable non-perturbative (NP) corrections.
A simple model for NP effects, already proposed by
Basham et al.\ in 1979 \cite{BBEL},
suggested that they should scale as $1/Q$, $Q$ being the total
annihilation energy (hardness scale).
A more detailed model for NP corrections to
the EEC was suggested by Collins and Soper in 1985
\cite{CS85}.  Operationally, they suggested modelling the NP effects
due to the transition from partons to hadrons as a kind of
smearing of the PT distribution.

In recent years power-suppressed NP contributions were studied for a
wide variety of hard cross sections~\cite{collection}.  In particular,
$1/Q$ contributions were predicted and phenomenologically quantified
for a number of jet shape observables such as thrust and jet masses,
the $C$-parameter and jet broadenings (for a review see~\cite{Vanc}).
Following the technique for analysing the power-behaved contributions
to hard observables developed in~\cite{DMW}, we are now in a position
to better understand the NP effects in the EEC distribution and to
relate the corresponding NP parameters with those emerging from the
analyses of jet shapes.  This is the main purpose of the present
article.

Let us start by recalling that
the energy-energy correlation (EEC) is defined as
\begin{equation}
\begin{split}
  \label{dsig}
\frac{d\Sigma(\chi)}{d\cos\chi}
&\equiv 
\frac{d\sigma}{\sigma\,d\cos\chi}
   \>=\>\sum_{n} \int \frac{d\sigma_n}{\sigma}\> \EEC(\chi)\>, \\ 
  \EEC(\chi) &= \sum_{a,b}\frac{ E_a\,E_b}{Q^2}\> 
  \delta(\cos\chi + \cos\Theta_{ab})\>,
\end{split}
\end{equation}
where the sums are over all final-state particles $a$ and $b$,
so that each pair of particles is counted twice. 
Here $\chi=\pi-\Theta_{ab}$ so that in the back-to-back region
$\chi\ll 1$.

Perturbatively, the correlation is dominated by the contribution from
the primary $\qq$ pair, while the $qg$ correlation produces a
subleading correction.

At the non-perturbative level, there are two physically different
confinement effects in the EEC. The first is an additional NP
contribution to the $\qq$ angular imbalance due to radiation of
secondary gluons with transverse momenta of the order of $\LQCD$,
which we call {\em gluers}~\cite{gluer}.  This contribution scales as
$1/Q^2$. A more important NP contribution comes from the correlation
between the (anti)quark and a {\em gluer}\/ with $E_g\sim\LQCD/\chi$,
which scales as $1/(Q\chi)$.  As a result of an interplay between the
NP and PT effects in the EEC distribution, the naive dimensions $1/Q$
and $1/Q^2$ get modified, in the back-to-back region, and the
power-suppressed contributions decrease with $Q$ much more slowly.

At the parton level, the quark-quark contribution to the integrated
distribution $\Sigma(\chi)$ for small values of $t=\tan(\half\chi)$
has the structure (to leading-logarithmic order)
$$ 
\Sigma_{\qq}(\chi) \>=\> f(\as\log^2t)\,.  
$$ 
The contribution from the (anti)quark--gluon correlation is one
log down: 
$$
\Sigma_{qg}(\chi) \>\sim\> \as\log\frac1t \cdot f(\as\log^2t)\,.  
$$
The single-logarithmic enhancement here comes from the collinear
singularity of the $qg$ matrix element at $t=0$.  However it can be
effectively absorbed into the $\qq$ contribution.  Indeed, adding the
energies of the quark and the gluon(s) {\em collinear}\/ with it
produces the initial quark energy, so that these two terms together
correspond to neglecting the quark energy loss in the $\qq$
correlation. Having performed the collinear subtraction, one is left
with the residual $qg$ contribution to $\Sigma$ at the level of a
correction of relative order $\as$.  Analogously, the correlation
between two secondary gluons starts at the $\as^2$ level and will be
neglected hereafter in the derivation of the resummed next-to-leading
PT distribution\footnote{The $\cO{\as^2}$ PT corrections coming from
  the $gg$ EEC, as well as from other sources, are taken care of by
  matching the approximate logarithmic distribution with the exact
  two-loop matrix element calculation, performed in
  Section~\ref{Matching}.}.

As a result, the EEC at small $\chi$, at the perturbative level, can
be simply treated by considering the correlation between the primary
quark and the antiquark, which are no longer aligned, because of
multiple gluon bremsstrahlung, but do not lose energy.

At the NP level, the leading power-behaved contribution due to the
quark-gluon correlation is proportional to
$$
\Sigma^{(\NP)}_{qg}(\chi) \propto \lrang{b}\cdot \Sigma_{\qq}^{(\PT)}(\chi) 
\,,
$$
where $\lrang{b}$, depending on $Q$ and the angle between the two
energy detectors, is the characteristic value of the impact parameter
determining the PT distribution.  In the back-to-back limit,
$\chi=\pi-\theta\to0$, one observes a power behaviour,
$\lrang{b}\propto Q^{-\gamma-1}$ with a non-integer anomalous
dimension $\gamma$ \cite{RW}.  As a result, the leading NP correction
to the height of the perturbative EEC plateau at $\chi=0$ becomes
$$
\frac{d\Sigma^{(\NP)}}{d\cos\chi}(\chi=0)\> 
\propto\> \frac{d\Sigma^{(\PT)}}{d\cos\chi}(\chi=0)\cdot Q^{-\gamma}\>. 
$$
The non-integer exponent depends on the treatment of the PT
coupling.  In particular, for the one-loop coupling with $n_f=3(5)$ we
obtain $\gamma\simeq 0.32(0.36)$
(see Appendix~\ref{App:oneloopqg}).

The final expression for the EEC which accounts for the leading power
effects has the following structure.  After extracting a kinematical
factor and the ``coefficient function'' factor $C=1+\cO{\as}$, we are
left with an expression based on the ``radiator'' $\cR(b)$, which
exponentiates in impact parameter space,
\begin{equation}
\begin{split}
\label{eq:EECdiff}
\frac{d\Sigma}{d\cos\chi} &= C(\as)\frac{(1+t^2)^3}{4} \> \cI(t) \>, \\
\cI(t) &= \frac{Q^2}{2} \int b\, db\>J_0(bQt)\>e^{-\cR(b)} 
\left( 1- 2b\scale \>+\> \cO{b^2\LQCD^2} \right).
\end{split}
\end{equation}
The linear NP correction $-2b\scale$ originates from the quark-gluer
correlation, where $\scale$ is a parameter (with the dimension of
mass) which characterises the NP interaction at small momentum scales,
\begin{equation}
    \scale= \frac{4C_F}{\pi^2}\cM\int_0^\infty dk\> \as^{(\NP)}(k)\>.
\end{equation}
The issue of the PT--NP
matching is explained in Sect.~\ref{App:NP-parameters}, and the
origin of the Milan factor $\cM$ is recalled in Appendix~\ref{App:NPqg}.

The radiator in \eqref{eq:EECdiff} contains its own NP component which
is {\em quadratic}\/ in $b$,
\begin{equation}
\begin{split}
  \cR(b) &= \cR^{(\PT)}(b) + \half b^2\cN\>, \\
\cN &= \frac{C_F}{2\pi}\int_0^\infty dk^2\>
\left(\ln\frac{Q^2}{k^2}-\frac12\right) \as^{(\NP)}(k)\>,
\end{split}
\end{equation}
Strictly speaking, this contribution should have been dropped since we
did not analyse a comparable quadratic effect which may come from the
$qg$ correlation. However we choose to keep the $\cN$ effect for two
reasons. Firstly, it allows us to verify that this quadratic NP term
affects the result much less that the leading $b\scale$ contribution.
Secondly, the $\cN$ contribution is logarithmically enhanced in $Q$,
which enhancement should not necessarily be present in the
next-to-leading power contribution from the $qg$ correlation.  A
complete analysis of $1/Q^2$ effects in the EEC remains to be done.

In the present paper we give a comparison of available data with
theoretical expectations based on ``default'' values of the relevant
NP parameters, without attempting a fit to extract the optimal values.
Our aim is to stimulate more detailed experimental studies of the EEC
in the back-to-back region, where there is a particularly interesting
interplay between perturbative and non-perturbative dynamics.

\section{Kinematics and resummation}
In this section, after introducing the energy-energy correlation and
the kinematics, we recall the relevant results of resummation in
the soft limit which are needed for power correction studies. To this
end, one needs to consider also the contributions coming from the
reconstruction of the running coupling at large distances. They are
obtained by using the dispersive method discussed in \cite{DMW}.  The
specific calculations for the EEC are similar to the ones performed
for shape variables (see \cite{DLMSThrust}). They are described in
detail in Appendix~\ref{App:radiator}.

\subsection{Kinematics in the soft limit}
At the parton level, the quantity EEC receives contributions from the
primary quark $p$ and antiquark $\bar{p}$ and the {\em secondary}\/
partons $k_i$.  In the soft limit the primary quark and antiquark
belong to opposite hemispheres. Neglecting the products of energies
$\om_i$ of the secondary partons, we have
\begin{equation}
\label{EEC2}
Q^2\cdot  \EEC(\chi) = 2E\bar{E}\delta(\cos\chi+\cos\Theta_{p\bar{p}}) 
  + 4\sum_{i=1}^n E\om_i \delta(\cos\chi+\cos\Theta_{pi})+\cO{\om_i\om_j}\>,
\end{equation}
where we have used the quark-antiquark symmetry of the matrix element.

For the parton momenta we use the Sudakov decomposition. Introducing
two light-light opposite vectors $P$ and $\bar P$, we write
\begin{equation}
  p = \zeta P\;,\qquad
  \bar{p} = \sigma P + \rho \bar{P} + p_t\;,\qquad
  k_i = \beta_i P + \alpha_i \bar{P} + k_{ti}\>,
\end{equation}
where we have taken $P$ along the quark direction and $2P\bar P=Q^2$.
In the soft limit all quantities
\[
\al_i,\>\be_i,\>1-\zeta,\>1-\rho,\>\frac{k_{ti}}{Q}\>,
\]
are small and of the same order, while $\sigma$ is much smaller
(quadratic in $p_t/Q$).  Neglecting quadratic soft terms, we have
\begin{equation}
 \label{EEC3}
 \begin{split}
 \EEC &\simeq \frac{(1+t^2)^2}{4}\left\{ 
 \zeta (\rho+\sigma) \delta\left(t^2-\frac{p_t^2}{Q^2\rho^2}\right)
 + 2\sum_i (\alpha_i+\beta_i)  
 \delta\left(t^2-\frac{k_{ti}^2}{Q^2\alpha_i^2}\right) \right\} \\
&= \frac{(1+t^2)^3}{4}\left\{\rho\,\zeta 
 \> \delta\left(t^2-\frac{p_t^2}{Q^2\rho^2}\right)
 + 2\sum_i \alpha_i \>\delta\left(t^2-\frac{k_{ti}^2}{Q^2\alpha_i^2}\right)
 \right\}.
 \end{split}
\end{equation}
Here we have used $\alpha_i\beta_i=k_{ti}^2/Q^2$,
$\>\sigma\rho=p_t^2/Q^2$ and
\[
t = \tan\frac{\pi-\theta}{2}\equiv\tan\frac{\chi}{2}\>, \qquad
\tan\frac{\pi-\theta_{p\bar p}}{2}= \frac{p_t}{Q\rho}\>, \quad 
\tan\frac{\pi- \theta_{pk_i}}{2}=\frac{k_{ti}}{Q\al_i}\>.
\]
The expression \eqref{EEC3} takes into account the recoil of the
quark-antiquark ($\rho\>\zeta \ne 1$ and $p_t\ne0$) against (soft)
secondary partons. It can be cast in a more transparent form by using
\begin{equation}
  \label{eq:recoiling}
 \rho\> \zeta= \left(1 -\sum_i \alpha_i\right)\left
(1 - \sigma - \sum_i \beta_i\right)
\>=\> 1-\sum_i\al_i-\sum_i\be_i + \ldots  
\end{equation}
where the dots correspond to terms quadratic in the soft scale.
Finally, using the fact that the matrix element is symmetric with
respect to exchange of $\al_i$ and $\be_i$, we can write (apart from
quadratic soft terms)
\begin{equation}
\label{EEC}
\EEC= \frac{(1+t^2)^3}{4}\left\{ 
\delta\left(t^2-\frac{p_t^2}{Q^2\rho^2}\right) + 2\sum_i\alpha_i  
\left[\, \delta\left(t^2-\frac{k_{ti}^2}{Q^2\alpha_i^2}\right)
- \delta\left(t^2-\frac{p_t^2}{Q^2\rho^2}\right) \right] \right\}.
\end{equation}
This form explicitly shows that the quantity EEC is infrared and
collinear safe.  In particular, it remains finite when a secondary
gluon happens to be collinear with the antiquark momentum, $\vtkti{i}
\to\>0$,
\begin{equation}
  \label{eq:vtktdef}
 \vtkti{i} \equiv \vec{k}_{ti}-\frac{\al_i}{\rho}\vec{p}_t\>,
\end{equation}
where the matrix element has collinear singularities (see
Appendix~\ref{App:R1}):
\begin{equation}
  \label{kappa}
\abs{M^2} \>\propto\> 
\left(\frac{2(pk_i)(k_i\bar p)}{(p\bar p)}\right)^{-1} 
\>=\> \frac1{\tkti{i}^2}\>\to\> \infty\>.
\end{equation}
Indeed, the first term on the right-hand side of \eqref{EEC} does not
depend on the secondary parton (gluon) variables. As a result,
collinear and soft divergences of the radiation probability cancel, in
the standard way, in the inclusive sum of real and virtual
contributions.  The second term is proportional to the secondary
parton momentum, $\al_i$, and therefore is present only in the real
contribution (quark-gluon correlation). Here the {\em soft}\/
singularity of the matrix element, $d\al_i/\al_i$, is damped by the
$\al_i$ factor, while the {\em collinear}\/ singularity, $\vtkti{i}
\to\>0$, is regularised by the vanishing difference of the delta
functions in the square brackets, the direct quark-gluon contribution
to the correlation and the subtraction term due to the antiquark
energy loss which was not included into the first term, see
\eqref{eq:recoiling}.

Hereafter we shall refer to the two terms in \eqref{EEC} as the $\qq$
and $qg$ contributions to the EEC, respectively.  Thus reorganised, the
$\qq$ contribution dominates the PT answer, while the $qg$ one gives
rise to the leading $1/Q$ NP correction.

\subsection{Resummation of soft contributions}
Resummation of multiple soft gluon radiation off the $\qq$ antenna is
necessary (and sufficient, in the leading order) for describing the
EEC in the back-to-back (small $t$) region.  In this approximation the
partial cross sections can be factorized as
\begin{equation}
  \label{eq:fact}
    \frac{d\sigma_n}{\sigma} = C(\as)\> dw_n\>. 
\end{equation}
Here $dw_n$ stands for the normalized $n$ soft parton emission
probability, and the ``coefficient function'' $C(\as)=1+\cO{\as(Q)}$
is included in order to match the soft-resummed expression with the
exact two-loop result.  According to \eqref{EEC}, the observable
\eqref{dsig} acquires two contributions
\begin{equation}
  \frac{d\sigma}{\sigma\,d\cos\chi} \>=\> 
 C(\as)\,\frac{(1+t^2)^3}{4}\> \cI(t)\>,\qquad 
\cI(t)= \cI_{\qq}(t) \>+\> \cI_{qg}(t)\>,
\end{equation}
where
\begin{eqnarray}
  \label{Iqq}
   \cI_{\qq}(t)&=& \sum_{n} \int dw_n\>
   \delta\left(t^2-\frac{\vec{p_t}^2}{Q^2\rho^2} \right)\>, \\
  \label{Iqg}
  \cI_{qg}(t)&=& 2\>\sum_{n} \int dw_n\>  \sum_i \alpha_i  
\left[\, \delta\left(t^2-\frac{k_{ti}^2}{Q^2\alpha_i^2}\right)
- \delta\left(t^2-\frac{p_t^2}{Q^2\rho^2}\right) \right] .
\end{eqnarray}
The distribution $\cI_{qg}(t)$ includes the recoiling part of the
$\qq$ contribution (second term in the square bracket) so that, as
observed before, the collinear singularities in $dw_n$ for
$\vtkti{i}\to 0$ are cancelled.
            
In Appendix~\ref{App:R1} we discuss in detail the soft parton emission
probabilities $dw_n$. They depend on the secondary parton momenta
$\vtkti{i}$ defined in \eqref{eq:vtktdef} and on the rescaled
antiquark momentum
\begin{equation}
  \label{eq:pdef}
 \vec{p} \>=\> \frac{\vec{p}_t}{\rho}
\end{equation}
and have the form\footnote{the $\{ \vtkti{i}, \al_i\}$ variables are
  convenient for describing partons in the {\em right}\/ hemisphere,
  i.e. the one opposite to the triggered quark, see below.}
\begin{equation}
  \label{dwn}
  dw_n=d^2p\>  \int\frac{d^2b}{(2\pi)^2} e^{i\vb\vtpt}\>
  dW_n\left(\{\vtkti{i},\al_i\};\vb\right)\>,
\end{equation}
where the distributions $dW_n$ are factorized in the momenta of the
secondary soft partons. To obtain such a factorization one needs to
introduce the integration over the impact parameter $\vb$ to represent
the transverse momentum conservation
\begin{equation}
  \label{deltakappa}
  \delta^2(\vec{p}_t+\sum_i\vec{k}_{ti})=
  \delta^2(\vec{p}+\sum_i\vtkti{i})=
  \int\frac{d^2b}{(2\pi)^2} e^{i\vb\vtpt}\>
  \prod_i\>e^{i\vb\vtkti{i}}\>.
\end{equation}
The factorization of $dW_n$ allows the soft parton resummation. In
particular one has
\begin{equation}
  \label{sumdW}
  \sum_n\int\>dW_n\left(\{\vtkti{i},\al_i\};\vb\right)\>=\>e^{-\cR(b)}, 
\qquad \cR(0)=0\>,
\end{equation}
with $\cR(b)$ the soft emission radiator.  

The distributions $dW_n$ are singular for $\tkti{i}\to 0$ and
$\al_i\to0$. At inclusive level, these singularities cancel against
corresponding singularities in the virtual contributions resummed by
Sudakov form factors included into $dW_n$.  As a result, the radiator
is collinear and infrared finite.

From \eqref{sumdW} we immediately obtain the $\qq$ contribution
$\cI_{\qq}(t)$ (see \eqref{Iqq})
\begin{equation}
\label{Iqqoriginal}
  \cI_{\qq}(t)= 
\frac{Q^2}{2} \>\int \>b db\>J_0(Qbt) \>e^{-\cR(b)}\>.
\end{equation}
Notice the normalization that in the limit of no secondary emission,
$\cR\to 0$, one has $ \cI_{\qq}(t)\to \delta(t^2)$.

The ``quark-gluon'' EEC, $\cI_{qg}(t)$, receives contributions from
each one of the secondary partons (see \eqref{Iqg}). Due to the
factorization of $dW_n$ the sum can again be expressed in terms of the
resummed distribution based on the standard radiator, with the
triggered parton singled out.  The details can be found in
Appendix~\ref{App:NPqg}.

\section{Soft emission radiator}

In this section we analyse the radiator, which contains both PT and NP
contributions.

The essential point is the reconstruction of the running coupling,
which requires a two-loop analysis.  To this accuracy the radiator is
given by the contributions of one and two soft partons and has the
form (see \cite{DLMSThrust})
\begin{equation}
  \label{cRoriginal}
  \cR(b)\>=\>\int d\om_1(k)\>[1-J_0(b\tkt)]\>+\>
  \int d\om_2(k_1k_2)\>[\,1-J_0(b|\vtkti{1}+\vtkti{2}|)\,]\>,
\end{equation}
where $d\om_1$ is the one ``real'' soft gluon emission distribution
with one-loop virtual correction included; $d\om_2$ is the two
non-independent ``real'' soft parton emission distribution.  The
precise expressions for $d\om_1$ and $d\om_2$ are recalled in
Appendix~\ref{App:R2}.  Notice that the last contribution is
inclusive, i.e.  the sum $\vtkti{1}+\vtkti{2}$ enters as argument of
the Bessel function.

The most natural way the running coupling appears in Minkowskian
observables \cite{GL} is through the dispersive relation,
\begin{equation}
  \label{aeff}
  \frac{\as(k)}{k^2}=\int_0^\infty \frac{dm^2}{(m^2+k^2)^2}\>\ae(m)\>,
\end{equation}
where the effective coupling~\cite{DMW}, $\ae(m)$, is the primitive
function of the discontinuity of $\as(m)$.  In the PT region, $m^2 \gg
\Lambda^2$, the effective coupling $\ae(m)$ differs from the standard
$\as(m)$ by $\cO{\as^3}$.  It is important to stress that the relation
\eqref{aeff} is supposed to be applicable both for large and small
momentum scales, and thus makes it possible to quantify the NP
contribution to the radiator.

By using the representation \eqref{aeff} we reconstruct the running
coupling in the radiator and obtain the following expression, see
Appendix~\ref{App:R2},
\begin{equation}
  \label{cR}
\cR(b) 
= \frac{C_F}{\pi} \int_0^{Q^2} dm^2\> \ae(m) \frac{-d}{dm^2}
 \int_0^{Q^2}\frac{d\tkt^2}{m^2+\tkt^2}
 \left[\, 1- J_0(b\tkt)\,\right] \ln\frac{Q^2e^{-\th}}{m^2+\tkt^2}\>.
\end{equation}
First we recall the PT result and then derive the leading NP part of
$\cR(b)$.

\subsection{PT part of the radiator}
By using \eqref{aeff} we show in Appendix~\ref{App:R2} that the PT
part of \eqref{cR} reproduces the well known next-to-leading
expression~\cite{2loop}
\begin{equation}
  \label{cRPT}
  \cR^{(\PT)}(b) =  \frac{C_F}{\pi} \int_0^{Q^2}\frac{d\tkt^2}{\tkt^2}
\> \as^{\PT}(\tkt) \left[\, 1- J_0(b\tkt)\,\right] 
\ln\frac{Q^2e^{-\th}}{\tkt^2}\>.
\end{equation}
Here the two-loop PT coupling $\as^{\PT}(\tkt)$ is taken in the
physical ``bremsstrahlung'' scheme, in which the coupling is defined
as the intensity of soft gluon radiation~\cite{CMW}.  Since the
observable is collinear and infrared finite, the $1/\tkt^2$
singularity is regularized by the factor $[1-J_0(b\tkt)]$.  In
\eqref{cRPT} we must keep $\tkt^2>\LQCD^2$.

The explicit expression for the PT radiator with the next-to-leading
accuracy was derived in \cite{Turnock}.  It is obtained by replacing
the factor $[1-J_0(b\tkt)]$ by the theta-function, see
Appendix~\ref{App:cRPT},
\begin{equation}
  \label{cRPT1}
   \cR^{(\PT)}(b) \>\simeq\>  \frac{C_F}{\pi} \int_0^{Q^2}
\frac{d\tkt^2}{\tkt^2}\> \as^{\PT}(\tkt) 
\ln\frac{Q^2e^{-\th}}{\tkt^2}\cdot\vartheta
\left(\tkt-\frac{2}{be^{\gamma_E}}\right). 
\end{equation}
This gives
\begin{equation}\label{eq:RPT}
\begin{split}
\cR^{(\PT)}(b) =& 
-\frac{16\pi C_F}{\beta_0^2}\Biggl[\frac{1}{\as}
\left(\ln(1-\ell)+\ell\right)
-\frac{3\beta_0}{8\pi}\ln(1-\ell)\\
&+\frac{\beta_1}{4\pi\beta_0}\left(\frac 1 2\ln^2(1-\ell)
+\frac{\ln(1-\ell)}{1-\ell}+\frac{\ell}{1-\ell}\right) \Biggr], 
\end{split}
\end{equation}
where
\begin{equation}
\ell =\beta_0\frac{\as}{2\pi}\ln\frac{bQe^{\gamma_E}}{2}\>,
 \qquad \beta_0=\frac{11N_c}{3}-\frac{2n_f}{3}\>,
 \quad\beta_1= 102-\frac{38n_f}{3}\>,
\end{equation}
and $\as$ means $\as(Q)$ in the bremsstrahlung scheme \cite{CMW}.  The
first line corresponds to the contribution from the one-loop running
coupling.  The radiators with the two-loop and one-loop $\as$ differ
at the level of an $\cO{\as^3\ln^3 b}$ term which is under control and
should be kept in the PT distributions.

The expression \eqref{eq:RPT} only makes sense for $0<\ell<1$, that
is, for $\bmin<b<\bmax$ where
\begin{equation}
\bmin = \frac{2}{Q}e^{-\gamma_E}\;,\qquad\qquad\bmax\>=\>\bmin
\exp\left(\frac{2\pi}{\beta_0\as}\right)\;,
\end{equation}
and therefore we define
\begin{subequations}
\begin{eqnarray}
\label{eq:bmaxcon}
\cR^{(\PT)}(b>\bmax) &=& \infty\;,\\
\label{eq:bmincon}      
\cR^{(\PT)}(b<\bmin) &=& 0\;.
\end{eqnarray}
\end{subequations}
From \eqref{Iqqoriginal} we find the PT part of the $\qq$ contribution
within single logarithmic accuracy in the soft limit,
\begin{equation}
  \label{IPT}
  \cI^{(\PT)}(t)\>\simeq\>\cI^{(\PT)}_{\qq}(t) = \frac{Q^2}{2}
  \int bdb\>J_0(Qbt)\>e^{-\cR^{(\PT)}(b)}\>.
\end{equation}
We recall that the $qg$ contribution does not contain single
logarithmic PT terms. The matching of the approximate resummed
expression \eqref{IPT} with the exact two-loop result will be dealt
with in Section~\ref{Matching}.

\subsection{NP part of the radiator\label{NPrad}}
The general expression \eqref{cR} also contains an NP contribution.
The latter is given by the non-analytic moments of $\delta\ae$, the NP
component of the effective coupling (see
Sect.~\ref{App:NP-parameters}).  According to \cite{DMW}, the leading
NP part of the radiator, $\delta\cR$, is obtained from \eqref{cR} by
replacing $\ae$ by $\delta\ae$ and extracting from the rest of
the integrand the leading term non-analytic in $m^2$ at $m^2=0$.  This
term comes from the region $\tkt^2\sim m^2\ll Q^2$ and therefore can
be obtained by expanding the Bessel function in \eqref{cR},
\begin{equation}
  \label{dcR}
\begin{split}
  \delta\cR(b)  &=  b^2\cdot\frac{C_F}{4\pi} 
  \int_0^\infty dm^2 \delta\ae(m) \frac{-d}{dm^2}
  \int_0^\infty \frac{\tkt^2\>d\tkt^2}{m^2+\tkt^2}\> 
  \ln\frac{Q^2e^{-\th}}{m^2+\tkt^2}\\
 &=  b^2\cdot\frac{C_F}{2\pi} \int_0^\infty dm^2\>\delta\ae(m)\>
 \ln^2\frac{Qe^{-\tq}}{m} \>\equiv\> \half b^2\cN.
\end{split}
\end{equation}
The upper limit in $m^2$ is irrelevant here since $\delta\ae$ has
support at small $m^2\sim \LQCD^2$.  In the second line we have
neglected terms which generate pieces analytic in $m^2$. The
quantity $\cN$ contains two NP parameters,
\begin{equation}
\label{cRNP}
\cN\>\equiv\>-A_{2,1}\left(\ln Q^2 -\th\right)+ \half A_{2,2}\>=\>
\cA_{2}\left(\ln Q^2 -\half\right) - \cA_2'\>. 
\end{equation}
Here $A$ and $\cA$ are the (log)moments of the NP effective coupling
$\delta\ae$ and of its dispersive companion $\as^{\NP}$, respectively
(see Sect.~\ref{App:NP-parameters} below).  From the relation between
$\cN$ and the NP component of the running coupling, $\as^{\NP}$,
\begin{equation}
\label{cN}
\cN \>=\> \frac{C_F}{2\pi} \int_0^\infty dk^2
\left( \ln\frac{Q^2}{k^2}-\frac12\right) \as^{\NP}(k)\>,
\end{equation}
it is clear that the answer remains invariant under the choice of the
scale of the logarithms in \eqref{cRNP}.

Taking account of the NP contribution to the radiator, the full
quark-quark EEC is
\begin{equation}
  \label{Iqqfull}
  \cI_{\qq}(t)= \frac{Q^2}{2}\int bdb\>J_0(Qbt)\>
e^{-\cR^{(\PT)}(b)}\> e^{-\half b^2\cN}\>.
\end{equation}
The NP effect in the quark-antiquark correlation is nothing but a
Gaussian smearing of the PT distribution $\cI^{(\PT)}(t)$. Indeed,
introducing a two-dimensional vector $\vec{t}$ we can represent the
answer in the form of a convolution
\begin{equation}
\label{Iqq1}
  \cI_{\qq}(t) = Q^2\>\int d^2t' \>
 \frac{e^{-(\vec{t}-\vec{t}')^2\frac{Q^2}{2\cN}}}{2\pi\,\cN}\>
\cI^{(\PT)}(t')\>,
\end{equation} 
with $\cI^{(\PT)}(t')$ the PT distribution given in \eqref{IPT}.

Equation \eqref{Iqq1} makes it possible to directly relate the NP
parameters entering into the definition of $\cN$ with observables
describing soft hadronization. Even with PT radiation switched off,
the direction of the leading quark undergoes a random walk in angle
due to formation of the NP hadronic plateau. As a consequence we
expect
\begin{equation}
  \cN = \lrang{k_\perp^2}\cdot n(Q)\>, \qquad 
  n(Q)= \rho_h\left(\ln Q^2 -\Delta\right)
\end{equation}
where $\lrang{k_\perp^2}$ is the value of the mean squared transverse
momentum of primary hadrons in jets, $\rho_h$ is the density of the
corresponding rapidity plateau and $\Delta$ the parameter determining
the effective length of the latter.  This analogy gives
\begin{equation}\label{plateau}
\begin{split} 
 \cA_2 = \lrang{k_\perp^2}\cdot \rho_h\>, \qquad
  \cA_2' = \cA_2\left(\Delta - \half\right).
\end{split}
\end{equation}
A naive estimate of these numbers, ignoring the effects of resonance
decays, may be obtained using the simplest exponential parametrization
of the transverse momentum distribution of soft hadrons,
\begin{equation}\label{kpspect}
P(k_\perp)\propto k_\perp\exp\left(-2k_\perp/\lrang{k_\perp}\right)
\qquad \lrang{k_\perp}\simeq 0.30-0.35\;\mbox{GeV}\;,
\end{equation}
together with the UA5 \cite{UA5} parametrization of the
charged multiplicity,
\begin{equation}\label{chmult}
\bar n_{ch} = 9.11\,s^{0.115}-9.50 \simeq 1.05\ln s - 0.39\;.
\end{equation}
Taking account of neutrals,
we find $\rho_p\simeq 1.5$ and $\Delta\simeq 0.4$,
while $\lrang{k_\perp^2} = \th\lrang{k_\perp}^2\simeq 0.13-0.18$ GeV$^2$,
so that we may expect the NP parameters to be
\begin{equation}\label{plateauest}
 \cA_2 \simeq 0.20 - 0.27\>, \qquad
  \cA_2'\simeq 0\;.                   
\end{equation}

\section{Quark-gluon correlation}
The quark-gluon distribution $\cI_{qg}(t)$ can be expressed in terms
of $d\om_1$ and $d\om_2$, the one- and two-soft parton distributions
which we have introduced for the radiator (see \eqref{cRoriginal}). We
have
\begin{equation}
\label{Iqgoriginal} 
\begin{split}
 \cI_{qg}(t)&=\int \frac{d^2\tpt d^2b} {(2\pi)^2} 
 \> e^{i\vec{b}\vtpt}\>e^{-\cR(b)}\> 
 \Biggl\{ \int d\om_1(k_1)\>u(k_1)\>e^{i\vec{b}\vtkti{1}}\\
 &+\>\int d\om_2(k_1k_2)\>[\,u(k_1)+u(k_2)\,]
   \>e^{i\vec{b}(\vtkti{1}+\vtkti{2})}  \Biggr\}, 
\end{split}
\end{equation}
where, according to \eqref{Iqg}, the functions $u(k_i)$ which probe
the EEC observable are
\begin{equation}
\label{uki}
u(k_i)\>=\> 2\al_i
\left[\delta\left(t^2-\frac{k^2_{t_i}}{Q^2\al_i^2}\right)
   -\delta\left(t^2-\frac{\tpt^2}{Q^2}\right)\right]\>.
\end{equation}
The relative transverse momentum $\vtkti{i}$ is defined in
\eqref{eq:vtktdef}. The distribution $d\om_1$ is singular in the
limit $\al_1\to0$ as well as when the gluon momentum becomes parallel
to that of the radiating quark, $\vtkti{1}\to0$.  The first (infrared)
singularity is compensated by the $\al_1$ factor in $u(k_1)$. The
collinear singularity cancels in the combination of delta functions in
\eqref{uki}. A similar regularisation occurs in $d\om_2$ with respect
to the ``parent gluon'' momentum. An additional (collinear)
singularity in $d\om_2$ when the two offspring partons become
parallel, $\vtkti{1}/\al_1=\vtkti{2}/\al_2$, gets absorbed into the
running coupling determining the emission of the parent gluon, see
Appendix~\ref{App:R2}.

\subsection{PT contribution}
As shown in Appendix~\ref{App:cRPTqg}, the PT component of
$\cI_{qg}(t)$ constitutes a small $\cO{\as}$ relative correction to
the ``quark-quark'' contribution
\[
\cI^{(\PT)}_{qg}(t) \sim \as(Q)\cdot\cI^{(\PT)}_{\qq}(t)\>.
\]
In the first two orders in $\as$ this contribution is fully taken into 
account by merging the approximate resummed expression with the exact 
$\cO{\as^2}$ result based on the matrix element calculation, as
will be explained below in Section~\ref{Matching}.

\subsection{NP contribution}\label{NPqg}
Hereafter we concentrate on the NP component $\cI_{qg}^{(\NP)}$ of the
quark-gluon correlation, which is the dominant power-behaved
contribution to the EEC.  Notice that the soft approximation which has
been used to derive \eqref{Iqgoriginal} suffices for this purpose.

Following the procedure introduced in \cite{DLMSThrust}, we compute in
Appendix~\ref{App:NPqg} the NP contribution and obtain
\begin{equation}
\cI_{qg}^{(\NP)}(t) 
  =\frac{C_F\cM}{\pi} \int_0^\infty \frac{dm^2}{m^2} \>
  \delta\ae(m)\cdot \delta\Omega(m^2)\>,
\end{equation}
where the leading non-analytic piece of the trigger function is
\begin{equation}
\label{eq:omexp}
  \delta\Omega(m^2) =\frac{2m}{Q} \int\frac{d^2t_g}{2\pi\, t_g^3} 
\Biggl(\cI_{\qq}(\abs{\vec{t}-\vec{t}_g\,}) - \cI_{\qq}({t})\Biggr),
\end{equation}
and $\cM$ the Milan factor.  
The NP contribution takes the form
\begin{equation}
\label{NP:Iqg}
\cI_{qg}^{(\NP)}(t) \>=\> \frac{2\scale}{Q} \int\frac{d^2t_g}{2\pi\, t_g^3} 
\Biggl(\cI_{\qq}(\abs{\vec{t}-\vec{t}_g\,}) - \cI_{\qq}({t})\Biggr),
\end{equation}
with $\cI_{\qq}(t)$ the PT distribution given in \eqref{IPT}.
The NP parameter $\scale$ can be related to the first moment of
the NP coupling defined Sect.~\ref{App:NP-parameters} below:
\begin{equation}
\label{NP:scaledef}
\scale = 2 A_{1,0}\cM = \frac 4 \pi \cA_1\cM\;.
\end{equation}
Using \eqref{IPT} and the relation
\begin{equation}
\label{eq:Besint}
\int\frac{d^2t_g}{2\pi\, t_g^3}\left(e^{i\vec{b}\vec{t}_g}-1\right)
= -b\;, 
\end{equation}
the result can be expressed in terms of the mean value of the impact
parameter $b$ averaged over the quark distribution as follows:
\begin{equation}
  \label{NPIqg}
 \cI_{qg}^{(\NP)}(t)=\frac{Q^2}{2} \int_0^\infty b\,db\>J_0(Qbt)\> 
e^{-\cR(b)}\>(-2b\scale)\>.
\end{equation}
Eq.~\eqref{NP:Iqg} has a clear physical interpretation.  It describes
the contribution to the EEC when one triggers on a gluer (a gluon with
$\kappa_t\sim m\sim\Lambda$) in a given direction, $\vec{t}$, with
respect to the thrust axis.  The corresponding direction of the
radiating quark is $\vec{t}_p=\vec{t}-\vec{t}_g$, where $\vec{t}_g$ is
the gluer direction with respect to the quark.  This contribution is
proportional to the gluer energy which, when expressed as the ratio
$\kappa_t/\theta$, produces in Eqs.~\eqref{eq:omexp} and
\eqref{NP:Iqg} an extra enhancement $1/t_g$ on top of the standard
logarithmic distribution $d^2t_g/t_g^2$.  It is this additional
singular factor which gives rise to the non-analytic contribution
$\sqrt{b^2}$ according to \eqref{eq:Besint}.

The convolution \eqref{NP:Iqg} remains finite due to ``real-virtual''
cancellation.  The subtraction term represents the quark energy loss
due to an unobserved gluer, which was disregarded in what we chose to
call the quark-quark EEC distribution. Note that one consequence of
this convenient subtraction convention is that what we call the
quark-gluon contribution is not positive definite.  We remark also
that the structure of the NP quark-gluon contribution $\cI_{qg}$ does
not suggest that it should be exponentiated.

Finally, observe that in the limit in which the accompanying
radiation is neglected, $\cR(b)\to 0$, one obtains
\begin{equation}
  \label{eq:1ord}
  \cI_{qg}^{(\NP)}(t)\>\to\> \frac{\scale}{Q\,t^3}\>, \qquad
  \frac{d\sigma^{(\NP)}}{\sigma\,d\cos\chi} 
\>\to\> \frac{2\scale}{Q} \left(\frac{1+t^2}{2t}\right)^3
                         =\frac{2\scale}{Q\,\sin^3\chi}\>,
\end{equation}
which is the first order dispersive result, in accord with 
the NP expectation of \cite{BBEL}.

By introducing the mean impact parameter $\lrang{b}=\lrang{b}(t,Q)$ we
can cast the NP $qg$ contribution \eqref{NPIqg} as
\begin{equation}
  \label{NPIqg2}
   \cI_{qg}^{(\NP)}(t)\>=\>
-2\lrang{b}\scale \cdot \cI_{\qq}(t) \>.
\end{equation}
For not too small values of $t$, such that $\as\log^2t<1$, we have
$\lrang{b}\sim 1/(tQ)$, which explains an additional $1/t$ enhancement
of the NP term on top of the kinematical $1/t^2$ factor in
\eqref{eq:1ord}. In the region $\as\log^2t>1$ the Sudakov suppression
effects slow down an increase of $\lrang{b}$ which flattens off and
tends to a $Q$-dependent constant in the $t\to0$ limit.
 
If we use the one-loop coupling in the (two-loop) PT radiator, this
behaviour can be explicitly computed (see Appendix~\ref{App:oneloop})
to yield a non-integer exponent, see \eqref{dIqg0},
\begin{equation}
  \label{eq:exponent}
 \lrang{b}(0,Q) \simeq \frac{1.0894}{\LQCD} 
\left(\frac{\LQCD}{Q}\right)^{0.3236}\>\mbox{for $n_f=3$}\;,
\qquad \simeq \frac{1.1356}{\LQCD} 
\left(\frac{\LQCD}{Q}\right)^{0.3595}\>\mbox{for $n_f=5$}.
\end{equation}
In Fig.~\ref{fig:meanb} this analytical prediction for $n_f=5$, shown
by the dashed line, is compared with result of a numerical integration
using the full two-loop perturbative radiator.  The two-loop curve
deviates only a little from the analytical one-loop calculation, which
is reassuring.

The same hadronization model that was used in the previous Section to
estimate the parameters $\cA_2$ and $\cA_2'$ gives for $\cA_1$ the
value 
\begin{equation}
\begin{split}\label{A1est}
 \cA_1\cM &= \frac{\pi}4 \rho_h\lrang{k_\perp} \simeq 0.34 - 0.40\;
\mbox{GeV}\;, 
\end{split}
\end{equation}
which follows from the comparison of the QCD and the ``tube model''
result for the leading power correction to the mean value of
thrust,~\cite{brwhadron}
\begin{equation}
Q\lrang{1-T}_{\mbox{\small NP}} = 2\scale \>=\> 2\rho_h
\lrang{k_\perp}\>
\simeq\> 1\;\mbox{GeV}\;.
\end{equation}

 \FIGURE[ht]
{
    \psfig{file=eec_meanb_fig.ps,width=0.7\textwidth}
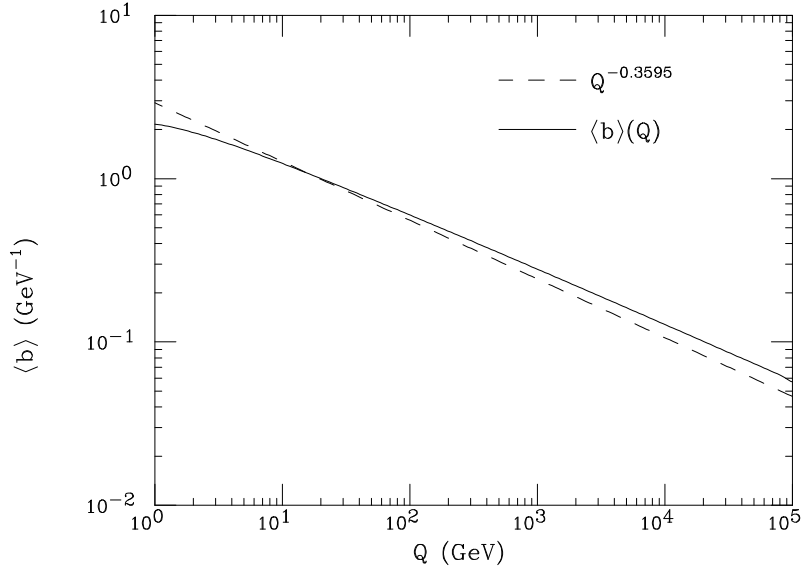
    \caption{Quasi-linear dependence $\log\lrang{b}(0,Q)$ on $\log Q$ with
      the expected slope}  
    \label{fig:meanb}
}

\section{Final results}
Combining the $\qq$ \eqref{Iqq} and $qg$ \eqref{NPIqg} contributions
we obtain
\begin{equation}
  \label{eq:insieme} \cI(t) \>=\> \cI_{\qq}+\cI_{qg} = \frac{Q^2}{2}
  \int_0^\infty b\, db\>J_0(Qbt)\> e^{-\cR^{(\PT)}(b)-\half
  b^2\cN}\left(1 - 2b\scale\right)\;,
\end{equation}
It should be clear that the NP $qg$ correlation gives the dominant $1/Q$
contribution, while the $\qq$ effect, at the level of $\log Q/Q^2$, 
is much smaller, both formally and numerically. 
In particular, we did not consider the next-to-leading NP
correction, potentially $\cO{Q^{-2}}$, coming from triggering $qg$.
However, it should still be legitimate to keep at least the leading
$\log Q$-enhanced piece in $\cN$, provided the subleading $1/Q^2$
correction from $qg$ is not $\log$-enhanced as well. To answer this
question one would have to analyse $\delta\Omega_{qg}$ further.

\subsection{Matching resummed and fixed-order predictions\label{Matching}}
We consider the integrated EEC distribution 
\begin{equation}
  \Sigma(\chi) \>=\>
\frac{1}{\stot}\int_0^{\chi}d\chi\,\frac{d\sigma}{d\chi}
\>=\> \int_0^t \frac{4t\,dt}{(1+t^2)^2}\> \frac{d\Sigma}{d\cos\chi} 
\end{equation}
and use \eqref{eq:EECdiff} to derive
\begin{multline}
\Sigma(\chi) = C(\as) \frac{Q^2}{2}\int bdb\>e^{-\cR(b)}(1-2b\scale)  
\int_0^t du\,u(1+u^{2})J_0(bQu) \\
= C(\as) \frac{tQ}{2} \int db\>e^{-\cR(b)}(1-2b\scale)
\left[\,  (1+t^2)J_1(bQt) - \frac{2t}{bQ} J_2(bQt)\,\right] .
\end{multline}
Neglecting corrections of the order of
$t^2\ll1$ in the back-to-back region, we finally arrive at
\begin{equation}
  \label{eq:Sigmaint}
\Sigma(\chi) \>=\> C(\as) \frac{tQ}{2} \int db\>J_1(bQt)\>e^{-\cR(b)}
(1-2b\scale)\>+\> \cO{t^2}\,.
\end{equation}
We now take advantage of the existing exact two-loop PT prediction for
EEC.  To this end we write
\begin{equation}
\Sigma(\chi) = \Sigma_{\mbox{\scriptsize resum}}(\chi)
 +\delta\Sigma(\chi)
\end{equation}
where $\Sigma_{\mbox{\scriptsize resum}}$ is the resummed prediction,
including NP corrections, and $\delta\Sigma$ is the matching
correction, which takes into account additional PT contributions up to
$\as^2$.

In order to obtain sensible predictions at small $\chi$, we have to be
careful to subtract and exponentiate {\em all}\/ logarithmic terms up
to this order, so that $\delta\Sigma$ remains finite as $\chi\to 0$.
The resummed expression based on the PT radiator accommodates all
logarithmically enhanced terms $\as^n\log^m t$ with $m\ge n$. The
finite non-logarithmic correction $\cO{\as}$ is taken care of in
\eqref{eq:Sigmaint} by the one-loop coefficient function \cite{CS85}
\begin{equation}
C(\as)= 
1-C_F\left(\frac{11}{2}
+\frac{\pi^2}{3}\right)\frac{\as}{2\pi}\>
\end{equation}
with  $\as = \as(Q)$ {\em in the $\MSbar$ renormalization scheme}. 
At the $\as^2$ level the first (and only) {\em singular}\/ subleading
logarithmic correction $\cO{\as^2\log t}$ appears which has not been
taken into account by the resummation procedure.  We therefore include
it into \eqref{eq:Sigmaint} to define
\begin{equation}\label{eq:Sigresum}
\Sigma_{\mbox{\scriptsize resum}}(\chi)\>=\>
C(\as)\frac{tQ}{2}\,
\exp\left[-G_{21}\tau\left(\frac{\as}{2\pi}\right)^2\right]
\int_0^\infty db\, J_1(b Q t)\> e^{-\cR(b)}(1-2b\scale)\; ,
\end{equation}
where
$t=\tan(\chi/2)$ and $\tau=\ln(1/t^2)$. The coefficient $G_{21}$ was
obtained by fitting the single-logarithmic term in the two-loop PT
contribution. In numerical evaluation of the integral in
Eq.~\eqref{eq:Sigresum}, the condition \eqref{eq:bmaxcon} was imposed,
so that impact parameters $b>\bmax$ do not contribute. We did not in
fact impose the condition \eqref{eq:bmincon} because its effect was
found to be negligible.

The matching correction $\delta\Sigma$ is then
\begin{multline}
\delta\Sigma(\chi) = \frac{1}{2}\Biggl[1
+\left(A_1(\chi)-B_{11}\tau-B_{12}\tau^2\right)\frac{\as}{2\pi}\\
+\left(A_2(\chi)-B_{21}\tau-B_{22}\tau^2-B_{23}\tau^3
-B_{24}\tau^4\right)\left(\frac{\as}{2\pi}\right)^2\Biggr]\;,
\end{multline}
where $A_1$ and $A_2$ are the one- and two-loop predictions,
obtained from the program EVENT2 \cite{event2}, and the $B_{ij}$'s are
the coefficients obtained by expanding Eq.~\eqref{eq:Sigresum}
to second order in $\as$, which gives\footnote{Note that
terms independent of $\chi$ are irrelevant to the differential
EEC and therefore we omit them.}
\begin{equation}\label{eq:Bijs}\begin{split}
B_{11}&= 3 C_F\nonumber\\
B_{12}&=  -C_F\nonumber\\
B_{21}&= -\left(\frac{33}{2}+\pi^2+4\zeta(3)\right)C_F^2
          +\left(\frac{67}{6}-\frac{\pi^2}{2}\right)C_F C_A
          -\frac{5}{3}C_F n_f-G_{21}\nonumber\\
B_{22}&=  \left(10+\frac{\pi^2}{3}\right) C_F^2
          +\left(\frac{\pi^2}{6}-\frac{35}{36}\right)C_F C_A
          +\frac{1}{18}C_F n_f\\
B_{23}&= -3 C_F^2-\frac{11}{9}C_F C_A +\frac{2}{9}C_F n_f\nonumber\\
B_{24}&=  \frac{1}{2} C_F^2\;.\nonumber
\end{split}\end{equation}
Requiring $\delta\Sigma(\chi)$ to be finite as $\chi\to 0$ then
gives $G_{21}\simeq 65$.

\subsection{Merging PT and NP contributions\label{App:NP-parameters}}
Within the dispersive method the analysis of the perturbative and 
non-perturbative contribution is performed by splitting
the coupling into two parts
\begin{subequations}
\begin{eqnarray}
\as(k) &=& \as^{\PT}(k) \>+\>\as^{\NP}(k)\>, \\
\ae(m) &=& \ae^{\PT}(m) + \delta\ae(m)\>.
\end{eqnarray} 
\end{subequations}
It is assumed that $\as^{\NP}(k)\>$
has a finite support, that is, it decreases fast at large $k^2$. 
This implies that $\ae(m)$ has only non-analytic $m^2$ moments,  
\begin{equation}
\label{Apq}
A_{2p,q}=\frac{C_F}{2\pi}\int 
\frac{dm^2}{m^2}\delta \ae(m)\>(m^2)^p\>\ln ^q m^2 \>,
\end{equation}
with $p$ non-integer or $q\neq 0$.
Using \eqref{aeff} it is straightforward to relate these parameters 
to the moments of $\as^{\NP}(k)\>$, 
\begin{equation}
 \label{eq:cApq}
\begin{split}
\cA_{2p}  &=   \frac{C_F}{2\pi} \int \frac{dk^2}{k^2} 
\>\left({k^2}\right)^p\> \as^{\NP}(k)\>, \\
\cA'_{2p}
  &=   \frac{d}{dp}\cA_{2p} 
   =   \frac{C_F}{2\pi} \int \frac{dk^2}{k^2}
\>\left({k^2}\right)^p\> \ln k^2 \> \as^{\NP}(k)\>,
\end{split}
\end{equation}
as follows \cite{DasWeb}
\begin{equation}
  \label{eq:7ext}
 A_{2p,q} \>=\> \frac{d^q}{dp^q}\left[\, \frac{\sin \pi p}{\pi p}\>\cA_{2p}
\,\right]. 
\end{equation}
In particular one has
\begin{equation}
  \label{eq:8}
  A_{1,0} =\frac{2}{\pi}\cA_1,\qquad
  A_{2,1} = -\cA_{2}\>, \qquad
  A_{2,2} = - 2\cA_2'+2\cA_2\>.
\end{equation}
In order to define these parameters more precisely, the problem of
merging the perturbative and non-perturbative contributions must be
addressed. The relevant procedure was discussed in detail in
\cite{DLMSuniv}.  It involves introducing an infrared matching scale
$\mI$ (typically chosen to be $\mI=$2 GeV), above which the NP
component of $\as$ is assumed to be negligible.  The PT prediction for
a given observable contains a contribution from the region $\mu<\mI$.
If it is calculated to next-to-leading order, then the PT coupling is
represented by its two-loop expansion with respect to $\as\equiv
\al_{\MSbar}(Q)$:
\begin{equation}
  \label{eq:aPTexp}
\as^{\PT}(k) = \as+\frac{\be_0}{2\pi}\left(\ln\frac{Q}{k} 
+\frac{K}{\be_0}\right)\as^2\;.
\end{equation}
The term proportional to $K$ accounts for mismatch between the
$\MSbar$ and bremsstrahlung renormalization schemes, with $K$ given
below in \eqref{Kdef}.

Defining the moments of the coupling on the interval $0<k<\mI$,
normalized in such a way that they would all be equal if $\as(k)$
were constant in this region,
\begin{equation}
  \label{eq:asmoms}
\bar\al_{p,q}(\mI) \equiv \frac{(p+1)^{q+1}}{q!\,\mI^{p+1}}
\int_0^{\mI} dk\,k^p\ln^q\frac{\mI}{k}\,\as(k)\;,
\end{equation}
we have
\begin{equation}
  \label{eq:aPTexp2}
\bar\al_{p,q}^{\PT}(\mI) = \as+\frac{\be_0}{2\pi}\left(\ln\frac{Q}{\mI} 
+\frac{K}{\be_0}+\frac{q+1}{p+1}\right)\as^2\;.
\end{equation}
By subtraction, we can now express the non-perturbative parameters
\eqref{eq:cApq} in terms of the full moments \eqref{eq:asmoms}.
In particular we have
\begin{equation}
  \label{eq:A2pexp}
\begin{split}
\cA_{2p} &= \mI^{2p}\cdot \frac{C_F}{2\pi p}
\left[\bar\al_{2p-1}(\mI)-\bar\al_{2p-1}^{\PT}(\mI)\right]\>, \\
\cA'_{2p} &= \mI^{2p}\cdot \frac{C_F}{2\pi p^2}
\left\{p\ln\mI^2
\left[\bar\al_{2p-1}(\mI)-\bar\al_{2p-1}^{\PT}(\mI)\right]
-\bar\al_{2p-1,1}(\mI)+\bar\al_{2p-1,1}^{\PT}(\mI)\right\} ,
\end{split}
\end{equation}
where $\bar\al_{2p-1}\equiv\bar\al_{2p-1,0}$.
Note that these quantities depend, via \eqref{eq:aPTexp}, on the order
of perturbation theory used to make the PT 
prediction.  If this is extended to next-to-next-to-leading order then
a further term of order $\as^3$, which can easily be computed, should
be added to \eqref{eq:aPTexp}.  
The corresponding PT terms in \eqref{eq:aPTexp2}, which diverge
factorially in higher orders, represent the start of the series
responsible for subtracting off the infrared renormalon divergence in
the {\em perturbative}\/ contribution to the observable.  Thus there
is no renormalon ambiguity in the sum of the PT and NP contributions.
 
The NP parameters $\scale$ and $\cN$ introduced above are now given by
\begin{subequations}
  \label{eq:cPcNfin}
\begin{eqnarray}
 \label{eq:lamfin}
\scale &=& \cM\frac{4C_F}{\pi^2}\mI
\left[\bar\al_0(\mI)-\bar\al_0^{\PT}(\mI)\right]\>, \\
\label{eq:cNfin}
\cN &=& \frac{C_F}{2\pi}\mI^2
\left\{\left(\ln\frac{Q^2}{\mI^2}-\frac 1 2\right)
\left[\bar\al_1(\mI)-\bar\al_1^{\PT}(\mI)\right]
+\bar\al_{1,1}(\mI)-\bar\al_{1,1}^{\PT}(\mI)\right\}\>.
\end{eqnarray}
\end{subequations}
Here $\cM$ in \eqref{eq:lamfin} is the Milan factor resulting from
the two-loop analysis discussed in Appendix~\ref{App:NPqg} (see
also~\cite{DLMSThrust}).  This factor is universal for all $1/Q$ jet
observables considered in $e^+e^-$ annihilation~\cite{DLMSuniv} and
DIS processes~\cite{DISmilan} and reads
\begin{equation}\label{cMdef}
\begin{split}
\cM \>=\>
1+{\be_0^{-1}}\left( 2.437C_A\>-\> 0.052n_f\right) 
\>=\> 1.920\>(1.665) \quad \mbox{for}\>\> n_f=5\>(0)\>.
\end{split}
\end{equation}

\section{Comparison with experiment\label{Comparison}}
In this section we compare the above predictions with
experimental data on the EEC near the backward direction.
At present the data are not plentiful and are not usually
binned in the optimal way for such comparisons. Therefore,
rather that attempting a detailed fit, we used ``default''
values of the relevant parameters. The only perturbative parameter
is the QCD scale, which we fixed to be\footnote{We consistently used
$n_f=5$ in the calculation of the radiator and the matching correction
$\delta\Sigma$, as well as in the Milan factor where it appears more
questionable.} 
\begin{equation}\label{eq:lmsnum}
\lms^{(n_f=5)} = 0.23\;\mbox{GeV}\;,
\end{equation}
which corresponds to $\as(M_Z)=0.118$. The three non-perturbative parameters
are moments of the coupling $\as(k)$ over the infrared region $0<k<\mI$,
which enter into Eqs.~\eqref{eq:cPcNfin} and are defined by \eqref{eq:asmoms}.
Choosing $\mI=2$ GeV, for the first two we take the values
\begin{equation}\label{eq:asmom01}
\bar\al_0(2\;\mbox{GeV}) = 0.50\;,\qquad\bar\al_1(2\;\mbox{GeV}) = 0.45\;,
\end{equation}
which come from analyses of $1/Q$ effects in event shapes \cite{A1num,Vanc}
and  $1/Q^2$ corrections to deep inelastic structure functions \cite{A2num},
respectively.  For the second log-moment, a new parameter which has not
been probed in other observables, we take the value according
to the model of Ref.~\cite{BRWmod}, which is also consistent with the
values \eqref{eq:asmom01}:
\begin{equation}\label{eq:asmom11}
\bar\al_{1,1}(2\;\mbox{GeV}) = 0.55\;.
\end{equation}
In terms of the dimensionful parameters defined in \eqref{eq:cApq},
these values correspond to
\begin{equation}\label{eq:cAnum}
\cA_1\cM\simeq 0.33\;\mbox{GeV}\,,\qquad\cA_2\simeq 0.2\;\mbox{GeV}^2
\,,\qquad\cA_2'\simeq 0.0\;\mbox{GeV}^2
\end{equation}
at $Q\sim M_Z$. Owing to the residual $Q$-dependence in Eq.~\eqref{eq:aPTexp2},
$\cA_1$ and $\cA_2$ are somewhat reduced at lower energies (falling to
$0.2$ and $0.12$ respectively
at $Q\simeq 10$ GeV), while $\cA_2'$ remains consistent with zero.

\FIGURE[ht]{
    \psfig{file=eec_low_fig.ps,width=0.9\textwidth}
    \caption{PLUTO \cite{pluto} data on the $\chi$ distribution of the EEC,
compared with PT and NP predictions.}
    \label{fig:low}
}

\FIGURE[ht]{
    \psfig{file=eec_zed_fig.ps,width=0.9\textwidth}
    \caption{SLD \cite{sld} and OPAL \cite{opal} data on the $\chi$
distribution of the EEC, compared with PT and NP predictions.}
\label{fig:zed} 
}

The theoretical predictions are compared with data on the distribution
in the angle $\chi=\pi-\theta$ at a range of energies in Figs.~\ref{fig:low}
and \ref{fig:zed}. The dot-dashed curves show the second-order PT
predictions, while the long-dashed curves display the results of purely
perturbative resummation. The short-dashed curves include the NP
quark-antiquark smearing effects, and the final results including
the NP quark-gluon correlation are shown by the solid curves.

The effect of PT resummation is to dramatically reduce the cross
section at small $\chi$, i.e.\ for nearly back-to-back kinematics
\cite{DDT,CS81,RW}, but not enough to match the data. The NP
contributions give a further reduction at small $\chi$ and an
enhancement at larger values. For $Q\sim M_Z$ the NP effects are
dominated by the quark-gluon contribution linear in $b$, while the
quadratic NP contributions to the radiator, due to quark-antiquark
smearing, become important at lower energies.

The distribution in $\chi$ has a kinematical suppression
of the most interesting region of small angles. The distribution in
$\cos\chi$, which is finite at $\chi=0$, is more informative.
Regrettably the only data set we could find that is binned in this
way is at the single energy $Q\simeq 30$ GeV \cite{pluto}. A comparison
with the theoretical expectations, again using the default parameters
given above, is shown in Fig.~\ref{fig:pluto}.

\FIGURE[ht]{
    \psfig{file=eec_pluto_fig.ps,width=0.9\textwidth}
    \caption{PLUTO \cite{pluto} data on the $\cos\theta$ distribution of the
EEC, compared with PT and NP predictions.}
    \label{fig:pluto}
}

All the predicted distributions are in reasonable agreement with the
experimental data. We would like to stress, however, that the
$\cos\theta$ distribution is much more sensitive to the NP effects.
With more precise data binned in $\cos\theta$ over a wide range
of energies, it should be possible to attempt quantitative fits to
extract the values of the important NP parameters, including the
new quantity $\bar\al_{1,1}$.

\section{Discussion}
In this paper we have investigated the leading power-behaved non-perturbative
contributions to the EEC. In particular we have demonstrated that the
power-suppressed contributions to the EEC distribution in the
back-to-back region are strongly modified by the interplay with purely
perturbative multiparton emission effects.
Thus the expected $1/Q$ behaviour of the leading non-perturbative term
due to the quark-gluon correlation turns into\footnote{The values of
  the exponents we present here correspond to $n_f=3$ and 5,
  respectively.  These estimates are based on an analytical treatment
  using the one-loop coupling. The actual two-loop exponent of the
  quark-gluon contribution, in particular, is even smaller at
  achievable energies, see Fig.~\ref{fig:meanb}.}
$(1/Q)^{0.32\mbox{-}0.36}$, while the $Q$-dependence of the
contribution due to NP smearing effects in the quark-quark EEC, $\log
Q/Q^2$, slows down to 
$(1/Q)^{0.58\mbox{-}0.65}$.
The latter effect should also be present in the differential
transverse momentum distribution of massive Drell-Yan lepton pairs in
hadron-hadron collisions, at small transverse momenta,
$p_\perp\ll M$.  Since the Drell-Yan process is fully inclusive
with respect to gluons, the ``$1/Q$'' effect which was leading in
the EEC case should be absent from the transverse momentum distribution,
as it is from the integrated cross section \cite{BBB,DMW}.

At the perturbative level, the present analysis includes the fully
resummed next-to-leading logarithmic expression for the EEC
distribution based on the two-loop radiator, which has been matched
with the exact order $\as^2$ result provided by EVENT2~\cite{event2}.

As far as non-perturbative physics is concerned, the aim of this
paper was to demonstrate consistency with the general framework
provided by the dispersive approach and with the concept of
universality of confinement effects. Therefore we have not
attempted a detailed quantitative analysis but rather have
compared with expectations based on other processes.

The leading NP effects are controlled by three phenomenological
parameters. The most important of them, which determines the NP
$qg$ contribution, is the one that describes $1/Q$ contributions
to the means and distributions of various jet shapes.  The value of
this parameter, $\bar\al_0(2\; \mbox{GeV})\simeq 0.50$, 
we have taken from jet shape phenomenology.

The other two parameters, $\bar\al_1$ and $\bar\al_{1,1}$, determine
the $\log Q$-enhanced and the constant terms of the NP ``$1/Q^2$''
$\qq$ contribution respectively.  The first of the two,
$\bar\al_1(2\;\mbox{GeV})\simeq 0.45$, we have taken from the
analysis of power corrections to the DIS structure functions.
Finally, the log-moment $\bar\al_{1,1}(2\;\mbox{GeV})\simeq 0.55$
we have borrowed from the model of Ref.~\cite{BRWmod}, since it is a
new quantity which has not yet been probed in other processes.

The results of these comparisons over a broad range of energies,
from 8 to 91 GeV (Figs.~\ref{fig:low}--\ref{fig:pluto}),
are encouraging. They show that the EEC distribution in
$\cos\theta$ in the back-to-back region, $\theta\sim\pi$,
is highly sensitive to NP effects, and therefore a fuller
experimental investigation of this region would be most welcome.

In Sects.~\ref{NPrad} and \ref{NPqg}
we pointed out the relation between these NP
parameters, or rather the $\cA$'s given by Eqs.~\eqref{eq:A2pexp},
and the characteristics of the rapidity plateau in ``soft'' hadron
production, Eqs.~\eqref{plateau} and \eqref{A1est}.
The standard values of the mean transverse momentum,
$\lrang{k_\perp}\simeq 0.3$ GeV,
and the number density, $\rho_h\simeq 1.5$,
give values of the NP parameters in reasonable agreement with
those obtained from other data and from the model of Ref.~\cite{BRWmod}.

The EEC in the back-to-back region has previously been studied
theoretically and phenomenologically by Collins and Soper \cite{CS81,CS85}.
As far as the perturbative aspects are concerned, what is new in
the present paper is the complete matching of resummed and 
fixed-order predictions, including exponentiation of all logarithmic
terms up to two-loop order.

Concerning the non-perturbative effects, Collins and Soper were the
first to point out the necessity of a leading NP contribution that is
linear in the ``impact parameter'' $b$. They also estimated the
coefficient of this contribution by fitting low-energy data.  In our
approach such a term arises inevitably from the quark-gluon
correlation and its magnitude is known from other observables, in
particular jet shapes. Contrary to the assumption of  Collins and Soper,
our approach does not suggest that such linear terms should be
exponentiated.

The dispersive approach also gives rise naturally to
contributions that are quadratic in $b$ (and $\log Q$ enhanced), which
can be interpreted as a NP smearing of the quark-antiquark
correlation. Collins and Soper's parametrization allowed for such
contributions but they were not included in their comparisons with
experiment.  In our treatment the linear and quadratic contributions
are comparable at low energies, with the former becoming dominant at
$Q\sim M_Z$. This emphasises again the importance of comprehensive
experimental studies over the widest possible range of energies.

\section*{Acknowledgements}

This research was supported in part by the EU Fourth Framework Programme
`Training and Mobility of Researchers', Network `Quantum
Chromodynamics and the Deep Structure of Elementary Particles',
contract FMRX-CT98-0194 (DG 12 - MIHT). GM and BRW are grateful for
the hospitality of the CERN Theory Division during a part of this
work.
We benefited from discussions with G.P.\ Korchemsky, 
G.P.\ Salam and D.E.\ Soper.

\appendix
\section{Radiator\label{App:radiator}}
\subsection{Phase space and momentum balance\label{App:PhSp}}
In terms of Sudakov variables, the phase space for the emission of 
the primary quark-antiquark pair together with $n$ partons  
\begin{equation}
  \label{dGn1}
  d\Phi_n= (2\pi)^4\delta^4\left(p+\bar{p}+\sum_i k_i -Q\right) 
           \frac{d^4p}{(2\pi)^3}\delta(p^2)
           \frac{d^4\bar{p}}{(2\pi)^3}\delta(\bar{p}^2)
           \prod_i\frac{d^4k_i}{(2\pi)^3}\delta(k_i^2)\>,
\end{equation}
can be written as
\begin{equation}
  \label{dGn2}
  d\Phi_n= \frac{d\Omega}{4\pi}
  \>{\zeta}\> 
  \frac{d\rho}{\rho} \>d^2p_t\>
  \delta^2(\vec{p}_t+\sum_i \vkti{i})\> \delta(1\!-\!\rho\!-\!\sum_i\al_i)
\prod_i 
\frac{d\al_i}{\al_i}\>\frac{d^2k_{ti}}{2(2\pi)^3}\>. 
\end{equation}
Introducing parton transverse momenta with respect to the antiquark
direction, $\vtkti{i}$ \eqref{eq:vtktdef}, and the angular antiquark
variable $\vtpt$ defined in \eqref{eq:pdef}, we can write
\begin{equation}
  \label{dGn3}
  d\Phi_n= \frac{d\Omega}{4\pi}
  \>d^2\tpt\> \delta^2(\vtpt+\sum_i \vtkti{i}) \> 
  \zeta\>\rho\>\prod_i \frac{d\al_i}{\al_i}\>
  \frac{d^2\tkti{i}}{2(2\pi)^3}\>, 
\end{equation}
with
\[
  \zeta=1-\sigma-\sum_i\be_i\>,\quad
  \rho=1-\sum_i\al_i\>, \quad 
  \sigma=\rho\>\frac{\tpt^2}{Q^2}.
\]
Since the soft matrix elements are factorized it is convenient to
express also the phase space in a factorized form.  This is obtained
by introducing the impact parameter $\vb$ to represent the transverse
momentum conservation
\begin{equation}
  \label{dGn4}
  d\Phi_n= \frac{d\Omega}{4\pi}
  \>d^2\tpt\> \int\frac{d^2b}{(2\pi)^2} 
  e^{i\vb\vtpt}\>
  \zeta\>\rho\>\prod_i \frac{d\al_i}{\al_i}\>
  \frac{d^2\tkti{i}}{2(2\pi)^3}\>e^{i\vb\vtkti{i}}\>.
\end{equation}
In the soft limit the upper bounds of the parton momentum integrations
can be arbitrarily chosen as $\tkti{i}<Q$ and $\al_i<1$.  An improper
treatment of the hard region of the phase space is then corrected by
introducing the coefficient function factor $C(\as)$ (see
\eqref{eq:fact}) and performing the matching of the approximate
resummed expression with the exact matrix element calculation to the
two-loop order, which was discussed in detail in Section~\ref{Matching}.

We will say that a secondary parton with
\[
\al_i>\be_i=\frac{k^2_{ti}}{\al_iQ^2}\>,
\qquad \mbox{or}\qquad
\be_i>\al_i=\frac{k^2_{ti}}{\be_iQ^2}\>,
\] 
is emitted in the {\em right}- or {\em left}-hemisphere respectively.
(Within this convention the quark belongs to the left hemisphere.)
We have chosen the Sudakov representation based on the {\em quark}\/
momentum direction. Therefore, as long as the quark and antiquark are generally
not back-to-back, the invariant phase space is not symmetric with
respect to left-right exchanges.

However, \eqref{dGn4} remains symmetric with respect to the R--L
hemispheres at the level of the terms {\em linear}\/ in gluon momenta,
which approximation is sufficient both for deriving the resummed PT
distribution and for extracting the leading NP effects.  Indeed, to
this accuracy we may write
\begin{equation}
\label{zetarho}
\begin{split}
\rho\>\zeta = \left(1-\sum_{i=1}^n\al_i\right) 
\left(1-\sum_{i=1}^n\be_i-\sigma\right)
\>\approx\> \prod_{i=1}^n (1-\al_i)\>(1-\be_i)\>,
\end{split}  
\end{equation}
where we have neglected the quadratic terms $\cO{\al_i\al_j}$ in the
first factor and both $\cO{\be_i\be_j}$ and $\sigma\propto p_t^2$ in
the second factor.
In conclusion, in the soft limit including quark recoil, we may use
\begin{equation}
  \label{dGn}
  \zeta\>\rho\>\prod_i \frac{d\al_i}{\al_i}\>\simeq\> 
  \prod_i 
  \left[\,d\al_i\>\frac{1-\al_i}{\al_i}\vartheta(\al_iQ-k_{ti})
        + d\be_i\>\frac{1-\be_i}{\be_i}\vartheta(\be_iQ-k_{ti})\,
  \right] .
\end{equation}
Here we have split the radiation into two hemispheres.  
For the emission in the right hemisphere, $\al_i>\be_i$,  
we included the factor $(1-\al_i)$ from \eqref{zetarho}, and 
similarly the factor $(1-\be_i)$ for the emission in the left hemisphere.

\subsection{One-loop radiator \label{App:R1}}
The soft multi-gluon radiation probability at one loop (multiple
independent soft gluon emission) takes the form
\begin{equation}
  \label{dMn}
\begin{split}
 dw_n &\>=\> d\Phi_n |M_n|^2 
  \simeq d^2\tpt\> \int\frac{d^2b}{(2\pi)^2} \>  e^{i\vb\vtpt}
  \\&
  \times \frac{1}{n!}\>\prod_i\left\{ \frac{C_F\as}{\pi} 
  \frac{d^2\tkti{i}}{\pi \tkti{i}^2}\>e^{i\vb\vtkti{i}}
  \left[\,d\al_i\>\frac{1-\al_i}{\al_i}\vartheta(\al_iQ-\tkti{i})
        + d\be_i\>\frac{1-\be_i}{\be_i}\vartheta(\be_iQ-\tkti{i})\,
  \right] \right\},
\end{split}
\end{equation}
where we have expressed the phase space for the emission in the right
and left hemisphere in terms of $\al_i$ and $\be_i$ respectively. 
Here we have substituted $\tkt$ for $k_t$ in the theta-functions
determining the Right-Left hemispheres. 
This pretty voluntary action is safe: the mismatch between the
lower limits of the $\al$-integration expressed in terms of the
transverse momentum defined with respect the quark, $k_t$, and the
antiquark, $\tkt$, is relatively small for small $k_t$ and/or small
$p_t$, which is our region of interest. 

The factor $(1-\al)/\al$ is the classical part of gluon emission
which, according to the celebrated Low-Barnett-Kroll theorem~\cite{LBK} 
embodies both the soft singularity, $d\al/\al$, and the first linear
correction, $d\al\cdot\cO{1}$.
Taking account of the true hard gluon radiation, $\half\al d\al$, this factor 
gets promoted to the full quark-gluon splitting function $P(\al)$, 
$$
\frac{1-\al}{\al}\>\>\Longrightarrow\>\> 
P(\al)\>=\> \frac{1+(1-\al)^2}{2\al}\>.
$$
Introducing the standard subtraction to accommodate virtual
contributions we arrive at the $n$-gluon emission probability, 
\begin{equation}
  dw_n\>=\>d^2\tpt\>\int \frac{d^2b}{(2\pi)^2}\> 
e^{i\vec{b}\vtpt}\>dW_n\>,
\end{equation}
where $dW_n$ factorizes (for given $\vb$ and $p\ll 1$) into $n_R$ and
$n_L$ soft gluons emitted into the right- and left-hemispheres,
\begin{equation}
  dW_n=dW_{n_R}\cdot dW_{n_L}\>, \quad n=n_R+n_L\>.
\end{equation}
Each of the two distributions is given by the soft factorized expression. 
The right-hemisphere distribution reads
\begin{equation}
  \label{dWnR}
\begin{split}
  dW_{n_R}\>&=\>
  exp\left\{ -\frac{C_F\as}{\pi}\int_0^1 d\al\>P(\al) 
  \int_0^{Q^2}\frac{d\tkt^2}{\tkt^2}\> \vartheta(\al Q-\tkt)\right\} 
\\ & \times
\frac1{n_R!}\prod_{i=1}^{n_R} 
\left\{ \frac{C_F\as}{\pi} d\al_i\>P(\al_i)\> 
\frac{d \tkti{i}^2\>\vartheta(\al_iQ- \tkti{i})}
{{\tkti{i}^2}} \> J_0(b\tkti{i})\right\}.
\end{split}
\end{equation}
A similar expression holds for the left-hemisphere contribution.
Summing over $n$  and integrating $dW_n$ one obtains
\begin{equation}
  \label{cRfirst}
\sum_n\int dW_n\>=\>e^{-\cR(b)}\>,
\end{equation}
where $\cR(b)$ is the one-loop radiator which receives contributions from
the radiation into both hemispheres. It is given by (see \eqref{dWnR})
\begin{equation}
  \label{cR1}
\begin{split}
\cR(b)&= \int_0^1 d\al\>\frac{C_F\as\>}{\pi}P(\al) 
\int_0^{Q^2}\frac{d\tkt^2}{\tkt^2}
\left[\,1-J_0(b\tkt)\,\right]\>  2\cdot \vartheta(\al Q-\tkt) \\
&\simeq \int_0^{Q^2}\frac{d\tkt^2}{\tkt^2}\> 
\frac{C_F\as\>}{\pi}
\>\left[\,1-J_0(b\tkt)\,\right]\>\ln\frac{Q^2e^{-3/2}}{\tkt^2}\>,
\end{split}
\end{equation}
where the factor $2$ in front of the theta function accounts for the
two hemispheres.  We have $\cR(0)=0$ which ensures the expected
normalization to the total cross section,
\begin{equation}
  \label{normaliz}
  \sum_n\int dw_n(k_1\cdots k_n,\vb)\>=\>
\int \frac{d^2p \>d^2b}{(2\pi)^2}\>e^{i\vb\vtpt}\>e^{-\cR(b)}\>=\>1\>.
\end{equation}

\subsection{Two-loop radiator \label{App:R2}}
We now consider the two-loop improvement.  By using the results of
\cite{DLMSThrust} we can generalise the form of the one-loop radiator
to include two-loop corrections in the soft region.  The radiator is
given by \eqref{cRoriginal} where the one and two uncorrelated soft
parton distributions $d\om_1$ and $d\om_2$ are given by
\begin{subequations}
\label{dom}
\begin{eqnarray}
\label{dom1}
  d\om_1(k)&\equiv& 4C_F\> d\al\> P(\al)\>\frac{d\tkt^2}{\tkt^2}\>
  2\vartheta(\al Q-\tkt)\left\{ \frac{\as(0)}{4\pi} + \chi(\tkt)\right\},\\
\label{dom2}
  d\om_2(k_1k_2)&\equiv& 4C_F\> d\Gamma_2(k_1,k_2)
  \left(\frac{\as}{4\pi}\right)^2 \,\frac1{2!} M^2(k_1,k_2) \>.
\end{eqnarray}
\end{subequations}
We briefly recall here the structure of these distributions~\cite{DLMSThrust}. 

The first distribution $d\om_1$ describes emission of a single real
gluon, with $\as(0)$ its on-shell coupling and $\chi(\tkt)$ the
one-loop virtual vertex correction given by
\begin{equation}
  \label{chi}
  \frac{\chi(\tkt)}{\tkt^2}= \int_0^{Q^2}
  \frac{d\mu^2}{\mu^2(\mu^2+\tkt^2)}
  \left(\frac{\as}{4\pi}\right)^2
  \left\{-2C_A\ln\frac{\tkt^2(\tkt^2+\mu^2)}{\mu^4}\right\}\>.
\end{equation}
The dispersive representation \eqref{chi} determines $\chi(\tkt)$ up
to a scheme-dependent constant, of order $\as^2$. Setting this
constant to zero, corresponds to the choice of the bremsstrahlung
scheme~\cite{CMW}.  In \eqref{chi} we have chosen to set the
ultraviolet integration limit at $\mu^2=Q^2$, rather than
$\mu^2=\infty$, in order to have the exact inclusive cancellation with
the real emission.  The error, $\cO{\as}$, induced by such a choice is
compensated by fixing the coefficient function appropriately.  The
theta function in \eqref{dom1} selects the gluons emitted in the
right-hemisphere; the accompanying factor of 2 takes care of the
contribution from the opposite (quark) hemisphere.

The distribution $d\om_2(k_1k_2)$ given in \eqref{dom2} corresponds to
the emission of two soft partons with 4-momenta $k_1$ and $k_2$.  The
uncorrelated ``Abelian'' two-gluon emission being subtracted off, the
rest can be described as the contribution from the radiation of a
(virtual) gluon followed by its decay into $\qq$ or $gg$ in the final
state. The corresponding matrix element $M$ can be found
in~\cite{DLMSThrust}.  The two-parton phase space $d\Gamma_2$ reads
\[
d\Gamma_2 \>=\> dm^2\,\frac{d\al}{\al}\,\frac{d^2\tkt}{\pi} 
\cdot dz\,\frac{d\phi}{2\pi}\>,
\]
where $\al=\al_1+\al_2$, $\vtkt=\vtkti{1}+\vtkti{2}$ and
$m^2=(k_1+k_2)^2$ are the ``parent gluon'' variables, while $z$ and
$\phi$ are the momentum fraction and relative azimuth of the two
secondary partons, $\qq$ or $gg$. 
The following kinematical relations hold:
\begin{equation}
\al_1=z\al\>,\>\> \al_2=(1-z)\al\>; \quad 
\vec{q_t}=\frac{\vtkti{1}}{z}-\frac{\vtkti{2}}{1-z}\>; \quad
m^2=z(1-z)q_t^2\>.
\end{equation}
Integrating $M^2$ over $z$ and $\phi$ one finds~\cite{DLMSThrust}
\begin{equation}
\label{M2int}
\int dz\>\frac{d\phi}{2\pi} \frac 1{2!} M^2(k_1,k_2) 
=\frac{1}{m^2(m^2+k_t^2)} \left(\frac{\as}{4\pi}\right)^2
\left\{2C_A\ln\frac{k_t^2(k_t^2+m^2)}{m^4}-\beta_0\right\},
\end{equation}
with $\be_0$ the one-loop coefficient of the beta function. 

Performing the $\al$-integration (over both hemispheres) of the
radiator $\cR(b)$ given in \eqref{cRoriginal} 
we arrive at
\begin{equation}
  \label{A:cR1}
\begin{split}
&\cR(b)= 4C_F \int_0^{Q^2}\frac{d\tkt^2}{\tkt^2}
\>\ln\frac{Q^2e^{-\th}}{\tkt^2}
\left(\frac{\as(0)}{4\pi} + \chi(\tkt) \right)
\left[\,1-J_0(b\tkt)\,\right]\>                    \\
 & + 4C_F \!\!\!\int_0^{Q^2}\!\!\frac{dm^2}{m^2}\left(\frac{\as}{4\pi}\right)^2
\!\!\!\int_0^{Q^2}\!\!\!\!\frac{d\tkt^2}{\tkt^2\!+\!m^2}
\ln\frac{Q^2e^{-\th}}{\tkt^2\!+\!m^2}
\left\{2C_A\ln\frac{\tkt^2(\tkt^2\!+\!m^2)}{m^4}\!-\!\beta_0\right\}
\left[1\!-\!J_0(b\tkt)\right].
\end{split}
\end{equation}
The collinear divergent terms, namely the logarithmic term on the
second line and the virtual contribution $\chi(\tkt)$, cancel if we
neglect the mismatch between the real and virtual phase space, which
is proportional to the factor $\ln(\tkt^2/(\tkt^2+m^2))$.  This
mismatch vanishes as $m^2/\tkt^2$ for $m^2\ll \tkt^2$ and, as we shall
see later, proves to be negligible within our accuracy both for the PT
and NP part of the radiator.

We are left then with the regular contribution proportional to
$\be_0$, and we get
\begin{equation}
  \label{A:cR2}
\begin{split}
\cR(b)= 4C_F \int_0^{Q^2} dm^2 
& \left\{\frac{\as(0)}{4\pi}\delta(m^2) 
-\frac{\be_0}{m^2}\left(\frac{\as}{4\pi}\right)^2\right\}
\\&\times
\int_0^{Q^2}\frac{d\tkt^2}{\tkt^2+m^2}\>\ln\frac{Q^2e^{-\th}}{\tkt^2+m^2}
\left[\,1-J_0(b\tkt)\,\right]\>.
\end{split}
\end{equation}
The two contributions in the curly brackets in \eqref{A:cR2} combine to
produce the running coupling. To see this we invoke the dispersive
representation \eqref{aeff} for $\as$ in terms of the effective
coupling $\ae$, which gives, under the $m^2$ integral,
\begin{equation}
\left\{ \frac{\as(0)}{4\pi}\,\delta(m^2)
-\frac{\be_0}{m^2}\,\left(\frac{\as}{4\pi}\right)^2\>
+\>\cO{\as^3} \right\}\cdot  \>=\> 
\frac{\ae(m)}{4\pi}\> \left(\frac{-d}{dm^2}\right)\cdot
\end{equation}
This leads to the following representation for the two-loop radiator
\eqref{A:cR2}:
\begin{equation}
  \label{A:cR}
\begin{split}
\cR(b) 
&= \frac{C_F}{\pi} \int_0^{Q^2} dm^2\> \ae(m) \frac{-d}{dm^2}
 \int_0^{Q^2}\frac{d\tkt^2}{m^2+\tkt^2}
 \left[\, 1- J_0(b\tkt)\,\right] \ln\frac{Q^2e^{-\th}}{m^2+\tkt^2}\\
&= \frac{C_F}{\pi} \int_0^{Q^2} dm^2 \ae(m)
 \int_0^{Q^2}\frac{d\tkt^2}{(m^2+\tkt^2)^2}\left[\, 1-J_0(b\tkt)\,\right]
 \left(\ln\frac{Q^2e^{-\th}}{m^2+\tkt^2}+1\right).  
\end{split}
\end{equation}

\paragraph{Perturbative equivalence.}
Within single logarithmic accuracy \eqref{A:cR} is perturbatively
equivalent to the well known expression
\begin{equation}
  \label{A:cRPT}
  \cR(b) =  \frac{C_F}{\pi} \int_0^{Q^2}\frac{d\tkt^2}{\tkt^2}
\> \as(\tkt) \left[\, 1- J_0(b\tkt)\,\right] 
\ln\frac{Q^2e^{-\th}}{\tkt^2}\>,
\end{equation}
with $\as$ taken in the bremsstrahlung scheme~\cite{CMW}.  

To show that \eqref{A:cR} and \eqref{A:cRPT} coincide at two-loop
level, we first extend the $m^2$-integration in \eqref{A:cR} to
infinity.  This produces a correction of the relative order
$\as(Q)\cdot k_t^2/Q^2$ which gives rise to a non-logarithmic
correction to $\cR$ of the order $\cO{\as(Q)}$, which we drop as
belonging to the coefficient function. Then from the dispersive
relation \eqref{aeff} one finds
\begin{equation}
\begin{split}
\int_0^{\infty} \frac{dm^2\>\ae(m)}{(m^2+k_t^2)^2} 
\left[\,\ln\frac{Q^2e^{-\th}}{m^2+k_t^2} +1\,\right] 
&=  \frac{\as(k_t^2)}{k_t^2} \ln\frac{Q^2e^{-\th}}{k_t^2}
+ \int_0^\infty \frac{dm^2\,[{\ae(m)}-{\as(k_t^2)}]}
{k_t^2\>(m^2+k_t^2)^2} \\
&=  \frac{\as(k_t^2)}{k_t^2} \ln\frac{Q^2e^{-\th}}{k_t^2}
\>+\>
\cO{\frac{\as^3(k_t^2)}{k_t^2}},
\end{split}
\end{equation}
which completes the proof.

\paragraph{Irrelevance of real-virtual mismatch.}
One more comment is warranted concerning a mismatch between the phase space
boundaries for the $\al$-integrations of the real and virtual
contributions leading to \eqref{A:cR1}.
Namely, the virtual integral extends down to
$\alpha\ge \tkt^2$, while the real one is cut off at $\alpha\ge
\tkt^2+m^2$. This tiny mismatch is nevertheless essential for the
power correction analysis.  Indeed, the difference of $\al$-integrals
of the real and virtual contributions produces
$$
 \ln\frac{Q^2e^{-\th}}{\tkt^2+m^2} -  \ln\frac{Q^2e^{-\th}}{\tkt^2}
= \ln\frac{\tkt^2}{\tkt^2+m^2}\>=\> -\frac{m^2}{\tkt^2}+ \ldots
\quad (\tkt^2\gg m^2)\>,
$$
which gives rise to a logarithmically enhanced $\cO{m^2}$ contribution 
coming from the integration region $m^2\ll \tkt^2 \ll b^{-2}$:
\begin{equation*}
   \int_0^{Q^2} \frac{d\tkt^2}{\tkt^2+m^2} \,
\ln\frac{\tkt^2(\tkt^2+m^2)}{m^4}\ln\frac{\tkt^2}{\tkt^2+m^2}
\left[\, 1- J_0(b\tkt)\,\right]
\>\simeq\>
-\frac{b^2}{2}\cdot m^2 \int_{m^2}^{b^{-2}}  \frac{d\tkt^2}{\tkt^2} 
\ln\frac{\tkt^2}{m^2} .
\end{equation*}
Taken at face value this would undermine the analysis of the quadratic
power correction to the $\qq$ EEC, since it seems to produce a
logarithmically enhanced term of the order of $m^2\ln^2m^2/Q^2$.
However, this non-cancellation occurs at the smallest kinematically
allowed values of $\al$, which correspond to the values of the
complementary Sudakov variable, $\beta$, at the edge of phase space
where $\be=1-\cO{m^2}$.
From the phase space \eqref{dGn4} there is a suppression factor
$1-\beta$ (coming from $\zeta$, see \eqref{zetarho}) which degrades a
contribution potentially non-analytic in $m^2$ down to $m^4$ at least.
Therefore we can neglect this mismatch and not to worry about the fact
that in this region neither \eqref{M2int} nor \eqref{chi}, which were
based on soft gluon approximation, are valid.

\section{Quark-gluon contributions\label{App:NPqg}}
By using the soft multi-parton distributions described in
Appendix~\ref{App:radiator}, we derive here the quark-gluon
distribution $\cI_{qg}(t)$ in the form given in \eqref{Iqgoriginal}.
We recall that $\cI_{qg}(t)$ is obtained by integrating the
partially inclusive quantity
\begin{subequations}
  \label{H}
\begin{eqnarray}
\label{H0}
H&=&\int d\om_1(k_1)\>u(k_1)\>e^{i\vec{b}\vtkti{1}}
\>+\>\int d\om_2(k_1k_2)\>[\,u(k_1)+u(k_2)\,]\>
e^{i\vec{b}(\vtkti{1}+\vtkti{2})},\\  
\label{H2}
u(k_i) &=& 2\al_i
\left[\delta(t^2-t^2_{k_i})-\delta(t^2-t^2_p)\right]\>; 
\quad\mbox{with}\quad 
t_p=\frac{\tpt}{Q}\>, \>\> t_{k_i}=\frac{k_{ti}}{Q\al_i}\>.
\end{eqnarray}
\end{subequations}
In order to make explicit the finiteness of $H$ and reconstruct the
running coupling,  we follow 
\cite{DLMSThrust}
and introduce an ``inclusive source'' 
(probing function) 
$u'(k)$ to write
\begin{equation}
\label{H1}\begin{split}
H&=\int d\om_1(k_1)\>u(k_1)\>e^{i\vec{b}\vtkti{1}}
\>+\>\int d\om_2(k_1k_2)\>u'(k)\>e^{i\vec{b}\vtkt^{\,\prime}}\\
&+\>\int d\om_2(k_1k_2)\>\left\{[\,u(k_1)+u(k_2)\,]\>e^{i\vec{b}\vtkt}
-u'(k)\>e^{i\vec{b}\vtkt^{\,\prime}}\right\}\>,
\end{split}
\end{equation}
where in the last two terms $k=k_1+k_2$, the momentum of the ``parent
gluon'' with positive virtuality (mass) $m$.  There is a freedom in
the choice of the expression for the probing function $u'(k)$ and
transverse momentum $\vtkt^{\,\prime}$ describing the contribution to
the EEC from a {\em massive}\/ object. The only requirement is that in
the $m^2\to0$ limit $u'(k)\exp(i\vec{b}\vtkt^{\,\prime})$ should
coincide with the standard massless parton contribution
$u(k)\exp(i\vec{b}\vtkt)$ given in \eqref{H2}.  This ensures that for
a collinear and infrared safe observable the difference vanishes for
$m^2\to0$, thus making the last term in \eqref{H1} finite in the limit
of collinear/soft parton splitting.  The first two terms of \eqref{H1}
remain finite to all orders, as in the case of the radiator, due to
the inclusive cancellation between the real and virtual corrections.

From \cite{DLMSThrust} we know that the two-loop analysis can be
greatly simplified if one defines $u'(k)$ by replacing the {\em
  transverse momentum}, $\tkt^2$, in the definition of the massless
source $u(k)$ by the {\em transverse mass}, $\tkt^{\prime
  2}=\tkt^2+m^2$.  Following this prescription we define
\begin{equation}
\label{A:uk}
\begin{split}
u'(k)\>&\equiv\> 2\al
\left[\,\delta(t^2 -t^{\prime2}_k)-\delta(t^2-t^2_p)\,\right]\>, \\
\tkt^{\prime2} &=\tkt^2+m^2\>,\quad 
k_t^{\prime2} = \left(\vtkt^{\,\prime} +\al\vtpt\,\right)^2\>,
\quad t'_k =\frac{k'_t}{\al Q}\>,
\end{split}
\end{equation}
with $\vtkt^{\,\prime}$ and $\vtkt$ parallel vectors. 
It is straightforward to verify that
the difference of the probing functions in the last term of 
\eqref{H1} vanishes for $m^2\to0$.

In Appendix~\ref{App:cRPTqg} we show that the PT component of
$\cI_{qg}(t)$ gives a next-to-next-to-leading contribution.
Here we discuss the NP component of $\cI_{qg}$ which provides the
dominant power correction to EEC.  We remark that it is legitimate to
use the soft approximation, upon which \eqref{Iqgoriginal} is based, to
analyse the $1/Q$ power contribution.

In \cite{DLMSThrust} it was proposed to group the terms of 
\eqref{H1} into three finite contributions,
\[
\cI_{qg}(t)\>=\>\cI_0+\cI_{in}+\cI_{ni}\>,
\]
the ``naive'', the ``inclusive'' and the ``non-inclusive'' contributions
we shall now discuss.

\paragraph{Naive contribution.}
In the first term, $\cI_0$, we reconstruct the running coupling
as in the case of the radiator. This is done by combining the $\as(0)$
term in $\int d\om_1(k)\>u(k)$ with the regular $\be_0$ part of the
parent gluon source,
$\int d\om_2(k_1k_2)\>u'(k)$, which emerges after integrating the
gluon decay probability over the offspring variables, see
\eqref{M2int}.  Similarly to the case of the radiator considered above,
we reconstruct the effective coupling by using \eqref{aeff}.  The
power contribution is then extracted by
substituting the NP 
component of the effective coupling, $\delta\ae$, for $\ae$.
One obtains
\begin{equation}
\label{A:Iqg0}
\begin{split}
\cI_0^{(\NP)}(t)=
&\frac{C_F}{\pi} \int_0^\infty dm^2 \delta \ae(m) \left(\frac{-d}{dm^2}\right) 
\int_0^\infty \frac{d^2\tkt}{\pi(\tkt^2+m^2)} \>\Om(\tkt^2+m^2)\>.
\end{split}
\end{equation}
Due to our choice of $k'_t$, the trigger function $\Om_0$ is a  
function of the transverse mass, $\tkt^2+m^2$, and is given by
\begin{equation}
\label{Om}
\Om(\tkt^2+m^2)\>=\>
\int \frac{d^2\tpt}{2\pi} \int bdb \>e^{-\cR(b)} J_0(b\tpt)
\> \int_0^1 2d\al 
\left[\delta(t^2-{t'}^2_{k})
-\delta(t^2-t^2_p)\right],
\end{equation}
where, in the leading (linear) approximation in $m\sim\tkt\sim\al Q$,
we have omitted the $b$-dependence in the inclusive source
\eqref{A:uk}.
Invoking \eqref{IPT} we arrive at
\begin{equation}
  \label{eq:Om2}
\Om(\tkt^2+m^2)\>=\> 4\int \frac{d^2t_{p}}{2\pi}\>\cI 
\left(t_p\right)\>\int_0^1 d\al \left[\delta\left(t^2-{t'}_k^2\right)
-\delta(t^2-t_p^2)\right]\>,  
\end{equation}
where we have invented the vector $\vec{t_p}=\vtpt/Q$.  Analogously
one can introduce the vector $\vec{t}_k^{\,\prime}$ with modulus
$t_{k}'$ and arbitrary direction.
Performing the integration over $\tkt^2$ in \eqref{A:Iqg0} by parts,
we obtain\footnote{Since only the terms non-analytic in $m^2$
  give non vanishing contributions to the $m^2$ integral with
  $\delta\ae$, the actual ultraviolet limit of the $\tkt$ integration, 
  $\tkt^2<{\tkt}_{\max}^2=\cO{Q^2}$, 
  does not matter.}
\begin{equation}
  \cI_0^{(\NP)}(t)\>=\> \frac{C_F}{\pi} \int_0^\infty dm^2 \delta \ae(m)\> 
\frac{\Om(m^2)}{m^2}\>.
\end{equation}
As a result of the $\tkt$ integration by parts we have
$\tkt=0$, and the relation \eqref{A:uk}, $\vec{t}_k^{\,\prime}=
\vec{\kappa}_t^{\,\prime}+\alpha\vec{p}$,
can be expressed as
\begin{equation}
\label{eq:tg}
\vec{t}_k^{\,\prime}= \vec{t}_g + \vec{t}_p\>, \quad
t_g=\frac{m}{\alpha Q}\>,
\end{equation}
the direction of $\vec{t}_g$ being arbitrary.
   
To make explicit the $m^2$ dependence of the trigger function $\Om(m^2)$ 
it is convenient to trade the $\alpha$ integration for that over $\vec{t}_g$
to arrive at
\begin{equation}
\label{rho}
  \Om(m^2) 
\>=\>\frac{m}{Q}\>\rho(t)\>, \qquad 
\rho(t)= 2\int\frac{d^2t_g}{2\pi t_g^3}\>
\left[\,\cI_{\qq}\left(\left|\vec{t}-\vec{t}_g\,\right|\right)-\cI_{\qq}(t)\,
\right].
\end{equation}
This equation has a clear physical meaning. 
The direction of the trigger is fixed to be $\vec{t}$. 
We have the standard logarithmic integration, $d^2t_g/t^2_g$, 
with $\vec{t}_g$ the direction of
the gluer with respect to the radiating quark, 
weighted by the perturbative distribution over the quark direction, 
$\vec{t}_p=\vec{t}-\vec{t}_g$.
The additional singular factor $1/t_g$ comes from weighting by the
gluer energy proportional to the ratio $m/t_g$, where $m\sim
\kappa_t\sim \Lambda$ is a typical finite transverse momentum scale
determining the leading power contribution.

We conclude with the expression
\begin{equation}
\cI_0^{(\NP)}(t)= \frac{2C_F}{\pi}\int dm \>\delta \ae(m)\cdot \rho(t)  
\>=\>\frac{2A_{1,0}}{Q}\cdot\rho(t) \>.
\end{equation}
(For definition of the NP
parameter $A_{1,0}$ see Sect.~\ref{App:NP-parameters}.)

\paragraph{Inclusive contribution.}
The ``inclusive'' contribution $\cI_{in}(t)$ is obtained by summing
the virtual correction $\chi$ in $\int d\om_1(k)\>u(k)$, see
\eqref{dom1}, together with the logarithmically divergent part of
$\int d\om_2(k_1k_2)\>u'(k)$, see \eqref{M2int}. Its NP part is
\begin{equation}
\label{A:Iqgin}
\begin{split}
\cI_{in}^{(\NP)}(t)&=
\frac{2C_AC_F}{\pi\be_0} \int_0^\infty dm^2\>\delta \ae(m) 
\left(\frac{-d}{dm^2}\right) 
\int_0^\infty \frac{d^2\tkt}{\pi(\tkt^2+m^2)} 
\>\ln\frac{\tkt^2(\tkt^2+m^2)}{m^4}
\>\Om_{in}\>,\\
\Om_{in}&\>\equiv\> \Om(\tkt^2+m^2)-\Om(\tkt^2)\>.
\end{split}
\end{equation}
The effective coupling here has been introduced by using the relation between
$\ae$ and the dispersive density of the coupling $\as$,
cf. \eqref{aeff}, 
\begin{equation}
  \label{eq:rhodef}
  \as(k) = - \int_0^\infty \frac{dm^2}{m^2+k^2}\> \rho(m)\>, \quad
  \rho(m) = \frac{d}{d\ln m^2}\, \ae(m)\>=\> 
-\be_0\frac{\as^2(m)}{4\pi}  + \ldots
\end{equation}
and by performing integration by parts.  

\paragraph{Non-inclusive contribution.}
Applying to the combination $\int
d\om_2(k_1k_2)\>[u(k_1)+u(k_2)-u'(k)]$ in \eqref{H1} the same
procedure of replacing $\as^2$ by the derivative of $\ae$, we
present the NP part of the ``non-inclusive'' contribution, $\cI_{ni}$,
in the form
\begin{equation}
\label{A:Iqgni}
\begin{split}
\cI_{ni}^{(\NP)}(t)&\>=\>
\frac{C_F}{\pi\be_0} \int dm^2\>\delta\ae(m) 
\left(\frac{-d}{dm^2}\right) 
\int \frac{d^2\tkt}{\pi} dz \frac{d\phi}{2\pi} \>M^2(k_1,k_2)\>\Om_{ni}\>,\\
\Om_{ni}&\>\equiv\> \Om(\tkti{1}^2)+\Om(\tkti{2}^2)-\Om(\tkt^2+m^2) \>.
\end{split}
\end{equation}

\paragraph{Milan factor.}
It is straightforward to verify that the combinations of trigger 
functions which enter into the inclusive and non-inclusive contributions 
are proportional, in the linear approximation, to the same function 
$\rho(t)$ defined by \eqref{rho}, which determines the naive contribution: 
\begin{eqnarray}
 \Om_{in} &=&\Om(\tkt^2+m^2) -\Om(m^2) 
 \simeq \frac{\rho(t)}{Q}\cdot \left(\sqrt{\tkt^2+m^2}-\tkt\right) , \\
  \Om_{ni} &=& \Om(\tkti{1}^2)+\Om(\tkti{2}^2)-\Om(\tkt^2)
\simeq\frac{\rho(t)}{Q}\cdot \left( \tkti{1}+\tkti{2} 
  -\sqrt{\tkt^2+m^2}\right). 
\end{eqnarray}
Such a structure is typical for the $1/Q$ power corrections to various 
jet shapes and leads to the {\em universal}\/ rescaling of the naive
contribution \eqref{A:Iqg0} 
by the so-called Milan factor; for details see~\cite{DLMSuniv}.

Taking account of the Milan factor, the full two-loop NP
component of the $qg$ contribution to the EEC becomes
\begin{equation}
\label{Milan}
  \cI_{qg}^{(\NP)}(t) = \cI_0^{(\NP)}(t)\cdot\cM
\>=\>\frac{\scale}{Q}\>\rho(t) \>,
\qquad \scale\equiv {2A_{1,0}\,\cM}\>.
\end{equation}

\section{Perturbative analysis of subleading corrections\label{App:PT}}

\subsection{Single-log corrections to the radiator \label{App:cRPT}}
In order to calculate $\cR^{(\PT)}$ with next-to-leading logarithmic
accuracy we introduce the primitive function
\begin{equation}
  \label{eq:10}
  \Phi(Q/\tkt) = \frac{2C_F}{\pi} 
  \int^Q_{\tkt}\frac{dk}{k} \as^{\PT}(k)
  \left(\ln\frac{Q^2}{k^2}-\frac 3 2\right)\>,
\end{equation}
where $\as^{\PT}$ is defined here in the bremsstrahlung scheme.
Writing $\as=\as^{\PT}(Q)$, $ \al=\as^{\PT}(k)$ for brevity,
we have to the required accuracy
\begin{equation}
 \ln\frac{Q^2}{k^2}  = \frac{4\pi}{\beta_0}
  \left(\frac{1}{\as}- \frac{1}{\al}
 + \frac{\beta_1}{4\pi\beta_0}\ln\frac{\as}{\al}\right),
\end{equation}
where
\begin{equation}
  \beta_0= 11-2n_f/3\>,\qquad\beta_1= 102-38n_f/3\>.
\end{equation}
Using the renormalization group equation, we can replace the $k$-integration
by one with respect to $\alpha$:
\begin{equation}\begin{split}
&  \Phi(Q/k) =\> \frac{16\pi C_F}{\beta_0} 
  \int_{\as}^{\al}\frac{d\al}{\al(\beta_0+\beta_1\al/4\pi)}
  \left(\frac{1}{\as}-\frac{1}{\al}+\frac{\beta_1}{4\pi\beta_0}
  \ln\frac{\as}{\al}-\frac{3\beta_0}{8\pi}\right)\nonumber\\
&= \> \frac{16\pi C_F}{\beta_0^2}\left[\frac{1}{\al}-\frac{1}{\as}
 +\frac{1}{\as}\ln\frac{\al}{\as} +\frac{\beta_1}{4\pi\beta_0}
 \left(\frac{\al}{\as}-1-\ln\frac{\al}{\as}
+\frac{1}{2}\ln^2\frac{\al}{\as}\right) 
-\frac{3\beta_0}{8\pi}\ln\frac{\al}{\as}\right]\;.\nonumber
\end{split}
\end{equation}
Introducing
\begin{equation}
\ell_0 =\beta_0\frac{\as}{2\pi}\ln\frac{Q}{k}\>,\nonumber
\end{equation}
we have, to the required accuracy,
\begin{equation}
\frac{1}{\al} = \frac{1}{\as}(1-\ell_0)
+\frac{\beta_1}{4\pi\beta_0}\ln(1-\ell_0)\>.
\end{equation}
Integrating by parts we have 
\begin{equation}
  \label{eq:11}
\begin{split}
 \cR^{(\PT)}(b)&=  \int_0^{Q} {d\tkt} 
\left[\,1-J_0(b\tkt)\,\right] \>\frac{-d}{d\tkt}\> \Phi(Q/\tkt) \\
&= \int_0^{bQ} dx\, J_1(x)\> \Phi(bQ/ x)\>+\>\cO{\as(Q)}\>, 
\qquad  x\equiv b\tkt\>,
\end{split}
\end{equation}
where we have neglected the non-logarithmic correction $\cO{\as(Q)}$
coming from the upper limit, which is taken care of by the coefficient
function.
Taking advantage of the fact that $\Phi$ is a slowly varying function, 
we may substitute into \eqref{eq:10} its logarithmic expansion,  
$$
 \Phi(bQ/ x) = \Phi(bQ) - \dot{\Phi}(bQ)\cdot \ln x
+ \cO{\as(b^{-1})}\>, \qquad \dot{f}(z)=\frac{d}{d\ln z}f\>,
$$
to obtain, with single-logarithmic accuracy,
\begin{equation}
  \label{eq:12}
\begin{split}
 \cR^{(\PT)}(b) &= \Phi(bQ) - \dot{\Phi}(bQ) \int_0^\infty dx\, J_1(x)
 \>\ln x +\ldots \\
&=   \Phi(bQ) - \dot{\Phi}(bQ) 
\left(\ln 2 -\gamma_E\right) + \ldots
\>=\> \Phi\left(\frac{bQ}{2}e^{\gamma_E}\right)+ \cO{\as}\,.
\end{split}
\end{equation}
Putting everything together, we find
\begin{equation}\label{eq:13cAPP}
\begin{split}
\cR^{(\PT)}(b) =& 
-\frac{16\pi C_F}{\beta_0^2}\Biggl[\frac{1}{\as}
\left(\ln(1-\ell)+\ell\right)
+\frac{\beta_1}{4\pi\beta_0}\left(\frac 1 2\ln^2(1-\ell)
+\frac{\ln(1-\ell)+\ell}{1-\ell}\right)\\
&-\frac{3\beta_0}{8\pi}\ln(1-\ell)
-\frac{1}{2\pi}\left(\beta_0\ln\frac{\mu}{Q}+K\right)
 \left(\ln(1-\ell)+\frac{\ell}{1-\ell}\right) \Biggr]
\end{split}
\end{equation}
where now
\begin{equation}
\label{eq:elldef}
\ell =\beta_0\frac{\as}{2\pi}\ln\frac{bQe^{\gamma_E}}{2}\>.
\end{equation}
Here we have allowed for an arbitrary renormalization scheme and
renormalization scale $\mu$, so that now $\as\equiv\as(\mu)$, in
both \eqref{eq:13cAPP} and \eqref{eq:elldef}. The last term in
Eq.~\eqref{eq:13cAPP} takes account of the scale and scheme dependence
to two-loop accuracy.  In the bremsstrahlung scheme $K=0$, whereas in
the $\MSbar$ scheme one has
\begin{equation}
\label{Kdef}
K =\frac{(67-3\pi^2)C_A-10 n_f}{18}\>.
\end{equation}

\subsection{Perturbative $qg$ contribution \label{App:cRPTqg}}
To show that the $qg$ correlation $\cI_{qg}(t)$ defined in
\eqref{Iqg} produces, at the PT level, 
a subleading $\cO{\as}$ correction to the EEC we consider the
``naive'' one-gluon contribution with the running coupling
reconstructed as explained above.  Measuring for brevity all momenta
in units of $Q$ we have
\begin{equation}
  \begin{split}
\cI_{qg}^{(\PT)}(t)&=2\int\frac{d^2b\,d^2\tpt}{(2\pi)^2}
e^{i\vb\vtpt}\> e^{-\cR(b)}
\int \frac{d^2\tkt\>e^{i\vb(\vtkt-\al\vtpt)}}{\pi(\vtkt-\al\vtpt)^2}
\>\frac{C_F\as(\tkt)}{\pi}\\
& \times \al\>d\al\,P(\al) 
\vartheta(\al-\tkt) 
\left[\, \delta\left(t^2-\frac{\tkt^2}{\al^2}\right) 
- \delta\left(t^2-\tpt^2\right)\,\right]\\
&= \frac{C_F\as}{\pi} \int_0^\infty b\,db\>  e^{-\cR(b)}
\int_0^{1} \frac{d\tkt^2\>d\tpt^2}{\abs{\tkt^2-\al^2\tpt^2}}
\int_0^1 \al\,d\al\>P(\al)\vartheta(\al-\tkt)\\
&\times J_0(b\tpt(1-\al))\>J_0(b\tkt)
\left[\, \delta\left(t^2-\frac{\tkt^2}{\al^2}\right) 
- \delta\left(t^2-\tpt^2\right)\,\right].
\end{split}
\end{equation}
Since, as we shall see shortly, the $\tkt$ integration is
non-logarithmic, we have chosen to neglect the running and pulled out
$\as$ as a constant factor.  Getting rid of the delta functions and
defining the common ``transverse momentum'' integration variable $q_t$
such that $q_t=p$ for the first delta function, and $q_t=\tkt/\al$ for
the second one, we arrive at
\begin{multline}
\cI_{qg}^{(\PT)}(t)\simeq \frac{C_F\as}{\pi} \int_0^\infty b\,db\>  e^{-\cR(b)}
\int_0^\infty \al\,d\al\>P(\al)\\
\times\int_0^{\infty}\frac{dq_t^2}{\abs{q_t^2-t^2}}\left[\,
J_0(bq_t(1-\al))J_0(bt\al)- J_0(bt(1-\al))J_0(bq_t\al)\,\right].
\end{multline}
Here we have substituted $\infty$ for the actual upper limit of the
$q_t$ integration, $q_t\la 1$ (i.e. $q_t\la Q$), because 
the region $q_t\ga 1$ corresponds to small impact parameters $b\le 1$ 
($b\le 1/Q$) and therefore produces a negligible contribution
$\cO{t^2}$ to the answer for $t\ll1$.

 The $\al$ and $q_t$ integrals converge and produce no logarithmic
 enhancement. For the sake of simplicity we shall demonstrate this
 property by considering the $qg$ contribution to the height of the
 ``PT plateau'', i.e. to the EEC distribution at $t=0$.  Setting $t=0$
 we have
\begin{equation}
  \begin{split}
 \cI_{qg}^{(\PT)}(0) &= \frac{C_F\as}{\pi}  \int_0^\infty b\,db\>  e^{-\cR(b)}
\int_0^1 \al\,d\al\>P(\al)
\int_0^\infty\frac{dq_t^2}{q_t^2}\left[\,
J_0(bq_t(1-\al))- J_0(bq_t\al)\,\right] \\
&= 2\frac{C_F\as}{\pi}\int_0^\infty b\,db\>  e^{-\cR(b)}
\int_0^1\,d\al\>(-\half\al)\cdot \ln\frac{\al}{1-\al} 
= -\frac{C_F\as}{2\pi}\int_0^\infty b\,db\>  e^{-\cR(b)}\>,
\end{split}
\end{equation}
where we have used  {\em antisymmetry}\/ with respect to
$\al\leftrightarrow(1-\al)$ to substitute  
$$
2\al P(\al)=1+(1-\al)^2= [\, 2-\al(1-\al)]-\al \>\Longrightarrow\> (-\al)\>, 
$$
and 
$$
\int_0^1\,d\al\>\al\ln\frac{\al}{1-\al}=  \frac12\>.
$$
We conclude that  $\cI_{qg}^{(\PT)}$ amounts to a non-logarithmic $\cO{\as}$
correction to the quark-quark contribution \eqref{Iqq}:
\begin{equation}
 \cI_{qg}^{(\PT)}(0) \simeq  \left[\, - \frac{C_F\as}{\pi}
 \right] \cdot \frac12\int_0^\infty b\,db\>  e^{-\cR(b)} 
\>=\>  \left[\, -\frac{C_F\as}{\pi} \right] \cdot \cI_{\qq}^{(\PT)}(0) \>.
\end{equation}

\section{Analytical estimates using one-loop coupling \label{App:oneloop}}
We recall that the PT part of the radiator to single logarithmic
accuracy is given by
\begin{equation}
  \label{PT:cR}
  \cR^{(\PT)}(b)=\frac{4C_F}{\pi}\int_{1/{\bb}}^{Q}\>
\frac{d\tkt}{\tkt}\>\as(\tkt)\ln\frac{Qe^{-\tq}}{\tkt}\>,
\quad \bb =\frac{be^{\gamma_E}}{2}\>,
\end{equation}
with $\as$ the two-loop coupling in the bremsstrahlung
scheme~\cite{CMW}.  In this Appendix we study the EEC distribution
neglecting $\beta_1$, the two-loop contribution to the running
coupling.  In so doing we lose control over single logarithmic
contributions to the PT radiator, starting from $\as^3\log^3$.  On the
other hand, this allows us to derive analytic expressions for the PT
distribution and for the NP corrections to it.  In particular we shall
derive in this approximation the non-integer exponents of the
$Q$-behaviour of the leading NP contributions to the EEC.

\subsection{Perturbative $\qq$ distribution \label{App:oneloopPT}}
The PT radiator with the one-loop $\as$ is
\begin{equation}
  \label{PT:cR1}
\cR^{(\PT)}(b) = -c 
\left[\,(L-\th)\ln\frac{L-\ell}{L}+\ell\,\right], 
\end{equation}
where
$$
L\equiv 2\ln\frac{Q}{\Lambda}, \quad 
\ell\equiv 2\ln (Q\bb)\>,
$$
and the numerical value of $c$ is
\begin{equation}
  \label{eq:cval}
c\equiv \frac{4C_F}{\be_0}\>=\frac{16}{33-2n_f}=0.5926\>,\quad 
\mbox{for}\>\> n_f=3\>.
\end{equation}
The PT evaluation only makes sense for $\ell < L$, which implies
\begin{equation}
\label{b0}
  b < b_0 \> \equiv \> \frac{2e^{-\gamma_E}}{\Lambda}\>.  
\end{equation}
The exponent of the radiator \eqref{PT:cR1} 
has the following Mellin representation,
\begin{equation}
  \label{PT:cR2}
e^{-\cR^{(\PT)}(b)}\>\vartheta(b_0-b) 
\>=\> \int_{a-i\infty}^{a+i\infty}
\frac{d\nu}{2\pi i}
\left(\frac{b_0}{b}\right)^{2\nu} \cdot \frac{F(L)}{c+\nu}
\left(\frac{c}{c+\nu}\right)^{c(L-\th)}\>,\quad a>0\>, 
\end{equation}
with
\begin{equation*}
 F(L) = e^{\th c} \left(\frac{e}{cL}\right)^{c(L-\th)}
\Gamma(1+c(L-\th))\>=\> \sqrt{2\pi cL}\> 
\left[\, 1+\cO{L^{-1}}\,\right].
\end{equation*}
The theta function on the left-hand side of \eqref{PT:cR2} is ensured
by the fact that the integrand  has no singularities in the
right half-plane, so that for $b>b_0$ the $\nu$-contour can be moved to
$a\to\infty$. 
Using 
\begin{equation}
\label{bes1}
 \int_0^{\infty} b\>db\>J_0(bt) \left(\frac{b_0}{b}\right)^{2\nu} 
= \frac{b_0^2}{2}\left(\frac{b_0t}{2}\right)^{2(\nu-1)}
\cdot \frac{\Gamma(1-\nu)}{\Gamma(\nu)}\>,
\end{equation}
we get 
\begin{subequations}
\label{cF}
\begin{eqnarray}
\cI_{\qq}(t)&=&\frac{Q^2}{2}\int_0^{b_0} b\>db\>J_0(Qbt) \>e^{-\cR^{(\PT)}(b)} 
\>=\> \int\frac{d\nu}{2\pi i}\> f(\nu,t)\>,   \\
\label{PT:fdef}
f(\nu,t)&=&\frac{F(L)}{t^2(c+\nu)}\frac{\Gamma(1-\nu)}{\Gamma(\nu)}
\left(\frac{\Lambda^2e^{\th}}{Q^2}\right)^{c\ln\frac{c+\nu}{c}}
\left(\frac{b_0Qt}{2}\right)^{2\nu}\>.
\end{eqnarray}
\end{subequations}
Here we have extended the $b$-integral to infinity since the
integrand represented by the right-hand side of \eqref{PT:cR2}
vanishes for $b>b_0$.  

The inverse Mellin transform \eqref{cF} can be formally evaluated by
closing the $\nu$-contour around the poles of $\Gamma(1-\nu)$ at
$\nu\!=\!1+p$, with $p=0,1,\ldots$ For $t=0$ only the pole at $\nu=1$
contributes and we derive the PT prediction for the height of the EEC
distribution in the back-to-back region, see \cite{RW},
\begin{equation}
\label{cF0}
  \cI_{\qq}(0)\>=\> 
  \frac{(b_0Q)^2\,F(L)}{4(c+1)} 
  \left(\frac{\Lambda^2e^{\th}}{Q^2}\right)^{c\ln\frac{c+1}{c}} 
\simeq \>\frac{e^{-2\gamma_E}}{c+1} \> 
\sqrt{2\pi c \ln\frac{Q^2}{\Lambda^2}} \cdot \frac{Q^2}{\Lambda^2}  
  \left(\frac{\Lambda^2e^{\th}}{Q^2}\right)^{c\ln\frac{c+1}{c}}
\!\!\!\!\!.
\end{equation}
Making use of this result  we can rewrite \eqref{PT:fdef} as
\begin{equation}
\label{effe'}
f(\nu,t)\>=\>\cI_{\qq}(0)\>\frac{\Gamma(1-\nu)}{\Gamma(\nu)}
\left(\frac{b_0Qt}{2}\right)^{2(\nu-1)}
\frac{c+1}{c+\nu}
\left(\frac{\Lambda^2e^{\th}}{Q^2}\right)^{c\ln\frac{c+\nu}{c+1}}
\end{equation}
For $t>0$ all the poles ($\nu=1+p$ with $p=0,1\ldots$) contribute, and
we obtain
\begin{equation}
\label{cFexp}
\cI_{\qq}(t)\>=\> \cI_{\qq}(0)\>
 \sum_{p=0}^\infty \frac{(-1)^p}{(p!)^2}
 \left(\frac{b_0 Qt}{2}\right)^{2p}
\frac{c+1}{c+1+p} 
 \left(\frac{\Lambda^2e^{\th}}{Q^2}\right)^{c\ln\frac{c+1+p}{c+1}}
\end{equation}
This expansion is convergent for any $(tQb_0)$.  However the series is
oscillating, and for large $(tQb_0)$ this representation is not
suitable for practical computation.

\subsection{Full $\qq$ distribution \label{App:oneloopqq}}
Taking account of the NP
contribution to the radiator, the full quark-quark EEC is given by 
\begin{equation}
  \label{Iqq2}
  \cI_{\qq}(t)= \frac{Q^2}{2}\int bdb\>J_0(Qbt)
\>e^{-\cR^{(\PT)}(b)}\> e^{-\half b^2\cN}\>=\>
 \int_{a-i\infty}^{a+i\infty} \frac{d\nu}{2\pi i}\>f(\nu,t)
\>\xi(\nu,t)\>,
\end{equation}
with 
\begin{equation}
\label{xi}
\begin{split}
\xi(\nu,t)\equiv & \int d^2t'\,  
\frac{e^{-t^{\prime2}/2\cN}}{2\pi\,\cN} 
\left(\frac{\vec{t}-\vec{t'}}{t}\right)^{2(\nu-1)}
\!\!\!\!\! =\int_0^{2\pi}\frac{d\phi}{2\pi}
\int_0^\infty \frac{dx^2}{x_0^2}e^{-{x^2}/{x_0^2}}
(1\!+\!x^2\!+\!2x\cos\phi)^{\nu-1} \\
&=\sum_{n=0}\frac{x_0^{2n}}{n!}\>
\left(\frac{\Gamma(\nu)}{\Gamma(\nu-n)}\right)^2; 
\qquad x_0^2\equiv\frac{2\cN}{t^2}\>.
\end{split}
\end{equation}
This function has no singularities in the finite $\nu$-plane. 
Taking the contributions from the poles in $f(\nu,t)$ we obtain 
the two equivalent expansions
\begin{equation}
\label{Iqqexp}
\begin{split}
\cI_{\qq}(t) 
&= \cI_{\qq}(0)\cdot \sum_{p=0}^\infty \frac{(-1)^p}{(p!)^2} 
\left(\frac{b_0Qt}{2}\right)^{2p}\frac{c+1}{c+1+p}
\left(\frac{\Lambda^2e^{\th}}{Q^2}\right)^{c\ln\frac{c+1+p}{c+1}}
X_p(t)\>,\\
&= \cI_{\qq}(0)\cdot \sum_{p=0}^\infty \frac{(-1)^p}{(p!)^2} 
\left(\frac{b_0Qt}{2}\right)^{2p}\frac{c+1}{c+1+p}
\left(\frac{\Lambda^2e^{\th}}{Q^2}\right)^{c\ln\frac{c+1+p}{c+1}}
Y_p\>,
\end{split}
\end{equation}
where
\begin{equation}
X_p(t)= \sum_{n=0}^p\frac{1}{n!}
\left(\frac{2\cN}{(tQ)^2}\right)^{n}
\left(\frac{p!}{(p-n)!}\right)^2 
\>=\>1\>+\>p^2\>\frac{2\cN}{(tQ)^2}\>+\ldots
\end{equation}
and
\begin{multline}
\label{Ydef}
Y_p\>=\> 
\sum_{n=0}^{\infty}\frac{(-\half b^2_0\cN)^n}{n!}
\frac{c+1+p}{c+1+p+n}
\left(\frac{\Lambda^2e^{\th}}{Q^2}\right)^{c\ln\frac{c+1+p+n}{c+1+p}} \\
=\> 1\>-\> \half b_0^2\cN\, \frac{c+1+p}{c+2+p}
\left(\frac{\Lambda^2e^{\th}}{Q^2}\right)^{c\ln\frac{c+2+p}{c+1+p}}
+\ldots
\end{multline}
In particular, $Y_0$ gives the height of the plateau at $t=0$ for the
full distribution, relative to the PT prediction \eqref{cF0}:
\begin{equation}
\label{Iqq0}
\cI_{\qq}(0)=
\cI_{\qq}(0) \cdot
 \sum_{n=0}^\infty   \frac{(-\half b_0^2\cN)^n}{n!}\>\frac{c+1}{c+1+n}\>
\left(\frac{\Lambda^2e^{\th}}{Q^2}\right)^{c\ln\frac{c+1+n}{c+1}}
\!\!\!\!.
\end{equation}
The exponents of successive power terms slowly
increase; their magnitudes oscillate and decrease factorially. 
For $n_f=3$ we have numerically 
\begin{equation}
\begin{split}
  \frac{\cI_{\qq}(0)}{\cI_{\qq}(0)}
& = 1- 
0.307\, (b_0^2\cN)
 \left(\frac{\Lambda^2e^{\th}}{Q^2}\right)^{0.289}
+ 
0.0554\,(b_0^2\cN)^2
 \left(\frac{\Lambda^2e^{\th}}{Q^2}\right)^{0.482}
+\ldots \\
&=
1-0.597\,\frac{\cN}{\Lambda^2}\left(\frac{\Lambda}{Q}\right)^{0.578}
+ 0.182 \left(\frac{\cN}{\Lambda^2}\right)^2 
      \left(\frac{\Lambda}{Q}\right)^{0.964} +\ldots
\end{split}
\end{equation}

\subsection{NP $qg$ contribution \label{App:oneloopqg}}
The NP part of $\cI_{qg}(t)$ is given by the convolution
\begin{equation}
\label{NP:qg}
  \delta\cI_{qg}(t) = \frac{2\scale}{Q}\>
\int \frac{d^2x}{2\pi x^3}\>[\,\cI_{\qq}(\vec{t}-\vec{x})-\cI_{\qq}(t)\,]
\>=\> \frac{2\scale}{tQ}
  \int\frac{d\nu}{2\pi i}\>f(\nu,t)\>\zeta(\nu)\>\>,
\end{equation}
with 
\begin{equation}
\label{zeta}
\begin{split}
  \zeta(\nu) &= \int_0^\infty \frac{dx}{x^2}\>\int_0^{2\pi}
\frac{d\psi}{2\pi}\>
[\,(1+x^2+2x\cos \psi)^{\nu-1}-1\,]\\
&= \int_0^1dx\int_0^{2\pi}\frac{d\phi}{2\pi}
\left\{ \frac{(1+x^2+2x\cos \psi)^{\nu-1}-1}{x^2}+
\frac{(1+x^2+2x\cos \psi)^{\nu-1}}{x^{2(\nu-1)}}-1
\right\}.
\end{split}
\end{equation}
The latter representation leads to the series expansion
\begin{equation}
  \label{zetaExp}
\zeta(\nu)\>=\>\sum_{k=0}\left(\frac{\Gamma(\nu)}{\Gamma(\nu-k)k!}\right)^2
\>\left(\frac{1}{2k-1}-\frac{1}{2(\nu-k)-3}\right).
\end{equation}
Observing that under the exchange $k\to \nu-k-1$ 
the first factor is symmetric and the second is antisymmetric, we
conclude that $\zeta(\nu)$ vanishes at the positive integer points 
$\nu=1+p$, $p=0,1,\ldots$, thus cancelling the poles of the PT 
function $f(\nu,t)$. 

The only remaining singularities of the $\nu$-integrand in \eqref{NP:qg} 
are the poles of $\zeta(\nu)$ at $\nu=\th+p$ with  $p=0,1, \ldots$.  
Evaluating the Mellin transform by closing the contour around these 
poles we get an expansion similar to that for the PT
distribution $\cI_{\qq}(t)$, 
\begin{equation}
  \label{dIqg2}
    \delta\cI_{qg}(t) = -
2\scale b_0\cdot \>\cI_{\qq}(0)\cdot
 \sum_{p=0}^\infty \frac{(-1)^p}{(p!)^2}
\left(\frac{b_0Qt}{2}\right)^{2p} \frac{c+1}{c+\th+p} 
 \left(\frac{\Lambda^2e^{\th}}{Q^2}\right)^{c\ln\frac{c+\th+p}{c+1}}\!\!\!\!.
\end{equation}
In particular, we immediately obtain for the leading NP correction
to the plateau height ($n_f=3$)
\begin{equation}
  \label{dIqg0}
\frac{\delta\cI_{qg}(0)}{
2\scale b_0 \>\cI_{\qq}(0)} 
= -\frac{c+1}{c+\th} 
  \left(\frac{\Lambda^2e^{\th}}{Q^2}\right)^{c\ln\frac{c+\th}{c+1}}
= -0.970\,  \left(\frac{\Lambda^2}{Q^2}\right)^{0.1618}\>.
\end{equation}

\subsection{Full EEC}
Putting together the  quark-quark and the 
quark-gluon contributions to the EEC we have
\begin{equation}
\begin{split}
\cI_{tot}(t)\>&\equiv\>\cI_{\qq}(t)+\cI_{qg}(t)
=  \int\frac{d\nu}{2\pi i}\>f(\nu,t)\>
\left\{\zeta(\nu)\>+\> 
\frac{2\scale} {tQ}\xi(\nu,t)\right\}\\
&= \cI_{\qq}(0)\cdot \sum_{p=0}^\infty \frac{(-1)^p}{(p!)^2} 
\left(\frac{b_0Qt}{2}\right)^{2p} \frac{c+1}{c+1+p}
 \left(\frac{\Lambda^2e^{\th}}{Q^2}\right)^{c\ln\frac{c+1+p}{c+1}}
\>Z_p\>,
\end{split}
\end{equation}
where
\begin{equation}
Z_p\> \equiv\>  1 \>-\> 2\scale b_0\>\frac{c+1+p}{c+\th+p} 
\left(\frac{\Lambda^2e^{\th}}{Q^2}\right)^{c\ln\frac{c+\th+p}{c+1+p}}
\>-\> \half b_0^2\cN\> \frac{c+1+p}{c+2+p}
\left(\frac{\Lambda^2e^{\th}}{Q^2}\right)^{c\ln\frac{c+2+p}{c+1+p}}
+ \> \ldots
\end{equation}
Setting $t=0$ we derive the leading NP power suppressed contributions
to the height of the plateau ($n_f=3$)
\begin{equation}
\begin{split}
\label{fine4}
\frac{\cI_{tot}(0)}{\cI_{\qq}(0)} &=
1-  1.57\> b_0\scale 
\,\left(\frac{\Lambda e^{\tq}}{Q}\right)^{0.323}
 - 0.307\,(b_0^2\cN) \left(\frac{\Lambda e^{\tq}}{Q}\right)^{1.16}
+\ldots \\
&= 1- 2.25\, \frac{\scale}{\Lambda}\left(\frac{\Lambda}{Q}\right)^{0.323} 
- 0.922\, \frac{\cN}{\Lambda^2} \left(\frac{\Lambda}{Q}\right)^{1.16}+\ldots 
\end{split}
\end{equation}
where $\cI_{\qq}(0)$ is the perturbative plateau in \eqref{cF0}.  The
first NP correction comes from the quark-gluon and the second from
the quark-quark correlation.

\end{document}